\documentclass[submission,copyright]{eptcs}
\usepackage[utf8]{inputenc}
\usepackage[T1]{fontenc}
\usepackage{alltt}
\usepackage{isabelle,isabellesym}
\usepackage{mathpartir}
\usepackage{subcaption}
\usepackage{amsfonts,amsthm}

\title{Hybrid Information Flow Analysis for Programs with Arrays}
\def\titlerunning{Hybrid Information Flow Analysis for Programs with Arrays}

\author{Gergö Barany\thanks{This work was supported by the French National
    Research Agency (ANR), project AnaStaSec, ANR-14-CE28-0014.}
\institute{CEA, LIST, Software Reliability Laboratory \\
           F-91191 Gif-sur-Yvette Cedex, France}
\email{gergo.barany@cea.fr}}
\def\authorrunning{Gergö Barany}

\newcounter{example}
\newenvironment{example}
{%
\refstepcounter{example}
\smallskip
\noindent \textbf{Example~\theexample.}
}
{
\hfill\(\Box\)%
\smallskip
}

\newcommand{\status}[1]{\ensuremath{\mathtt{\underline{#1}}}}

\isabellestyle{it}
\newcommand{\snip}[4]
  {\expandafter\newcommand\csname #1\endcsname{#4}}
\snip{instr}{1}{2}{%
\isacommand{datatype}\isamarkupfalse%
\ instr\ {\isacharequal}\ Skip\isanewline
\ \ \ \ \ \ \ \ \ \ \ \ \ \ \ \ \ \ \ \ \ \ \ \ \ \ {\isacharbar}\ Assign\ lval\ expr\ \ \ \ \ \ \ \ \ \ \ \ \ \ \ \ \ \ \ \ \ \ \ \ \ {\isacharparenleft}{\isachardoublequoteopen}{\isacharunderscore}\ {\isacharcolon}{\isacharcolon}{\isacharequal}\ {\isacharunderscore}{\isachardoublequoteclose}\ {\isadigit{5}}{\isadigit{2}}{\isacharparenright}\isanewline
\ \ \ \ \ \ \ \ \ \ \ \ \ \ \ \ \ \ \ \ \ \ \ \ \ \ {\isacharbar}\ AssignArrayElem\ lval\ expr\ \ \ \ \ \ \ {\isacharparenleft}{\isachardoublequoteopen}{\isacharunderscore}\ {\isacharcolon}{\isacharcolon}{\isacharequal}{\isacharprime}\ {\isacharunderscore}{\isachardoublequoteclose}\ {\isadigit{5}}{\isadigit{2}}{\isacharparenright}\isanewline
\ \ \ \ \ \ \ \ \ \ \ \ \ \ \ \ \ \ \ \ \ \ \ \ \ \ {\isacharbar}\ Seq\ instr\ instr\ \ \ \ \ \ \ \ \ \ \ \ \ \ \ \ \ \ \ \ \ \ \ \ \ \ \ \ {\isacharparenleft}\isakeyword{infixr}\ {\isachardoublequoteopen}{\isacharsemicolon}{\isacharsemicolon}{\isachardoublequoteclose}\ {\isadigit{5}}{\isadigit{1}}{\isacharparenright}\isanewline
\ \ \ \ \ \ \ \ \ \ \ \ \ \ \ \ \ \ \ \ \ \ \ \ \ \ {\isacharbar}\ If\ expr\ instr\ instr\isanewline
\ \ \ \ \ \ \ \ \ \ \ \ \ \ \ \ \ \ \ \ \ \ \ \ \ \ {\isacharbar}\ While\ expr\ instr%
}
\snip{updatedef}{1}{2}{%
\isacommand{fun}\isamarkupfalse%
\ collect{\isacharunderscore}updates\ {\isacharcolon}{\isacharcolon}\ {\isachardoublequoteopen}alias{\isacharunderscore}function\ {\isasymRightarrow}\ instr\ {\isasymRightarrow}\ block\ set{\isachardoublequoteclose}\ \isakeyword{where}\isanewline
{\isachardoublequoteopen}collect{\isacharunderscore}updates\ S\isactrlsub P\ Skip\ {\isacharequal}\ {\isacharbraceleft}{\isacharbraceright}{\isachardoublequoteclose}\ {\isacharbar}\isanewline
{\isachardoublequoteopen}collect{\isacharunderscore}updates\ S\isactrlsub P\ {\isacharparenleft}x\ {\isacharcolon}{\isacharcolon}{\isacharequal}\ {\isacharunderscore}{\isacharparenright}\ {\isacharequal}\ S\isactrlsub P\ x{\isachardoublequoteclose}\ {\isacharbar}\isanewline
{\isachardoublequoteopen}collect{\isacharunderscore}updates\ S\isactrlsub P\ {\isacharparenleft}x\ {\isacharcolon}{\isacharcolon}{\isacharequal}{\isacharprime}\ {\isacharunderscore}{\isacharparenright}\ {\isacharequal}\ S\isactrlsub P\ x{\isachardoublequoteclose}\ {\isacharbar}\isanewline
{\isachardoublequoteopen}collect{\isacharunderscore}updates\ S\isactrlsub P\ {\isacharparenleft}Seq\ i\isactrlsub {\isadigit{1}}\ i\isactrlsub {\isadigit{2}}{\isacharparenright}\ {\isacharequal}\ collect{\isacharunderscore}updates\ S\isactrlsub P\ i\isactrlsub {\isadigit{1}}\ {\isasymunion}\ collect{\isacharunderscore}updates\ S\isactrlsub P\ i\isactrlsub {\isadigit{2}}{\isachardoublequoteclose}\ {\isacharbar}\isanewline
{\isachardoublequoteopen}collect{\isacharunderscore}updates\ S\isactrlsub P\ {\isacharparenleft}If\ {\isacharunderscore}\ i\isactrlsub {\isadigit{1}}\ i\isactrlsub {\isadigit{2}}{\isacharparenright}\ {\isacharequal}\ collect{\isacharunderscore}updates\ S\isactrlsub P\ i\isactrlsub {\isadigit{1}}\ {\isasymunion}\ collect{\isacharunderscore}updates\ S\isactrlsub P\ i\isactrlsub {\isadigit{2}}{\isachardoublequoteclose}\ {\isacharbar}\isanewline
{\isachardoublequoteopen}collect{\isacharunderscore}updates\ S\isactrlsub P\ {\isacharparenleft}While\ {\isacharunderscore}\ i{\isacharparenright}\ {\isacharequal}\ collect{\isacharunderscore}updates\ S\isactrlsub P\ i{\isachardoublequoteclose}\isanewline
\isanewline
\isacommand{fun}\isamarkupfalse%
\ update\ {\isacharcolon}{\isacharcolon}\ {\isachardoublequoteopen}alias{\isacharunderscore}function\ {\isasymRightarrow}\ instr\ {\isasymRightarrow}\ label\ {\isasymRightarrow}\ label{\isacharunderscore}memory\ {\isasymRightarrow}\ label{\isacharunderscore}memory{\isachardoublequoteclose}\ \isakeyword{where}\isanewline
{\isachardoublequoteopen}update\ S\isactrlsub P\ prog\ s\ {\isasymGamma}\ {\isacharequal}\ {\isacharparenleft}{\isasymlambda}b{\isachardot}\ if\ b\ {\isasymin}\ {\isacharparenleft}collect{\isacharunderscore}updates\ S\isactrlsub P\ prog{\isacharparenright}\ then\ {\isasymGamma}{\isacharparenleft}b{\isacharparenright}\ {\isasymsqunion}\ s\ else\ {\isasymGamma}{\isacharparenleft}b{\isacharparenright}{\isacharparenright}{\isachardoublequoteclose}%
}
\snip{evalstmt}{1}{2}{%
\begin{isamarkuptext}%
\begin{center}
\isa{\mbox{}\inferrule{\mbox{}}{\mbox{E{\isacharcomma}\ S\isactrlsub P{\isacharcomma}\ pc\ {\isasymturnstile}\ Skip{\isacharcomma}\ M{\isacharcomma}\ {\isasymGamma}\ {\isasymRightarrow}\ M{\isacharcomma}\ {\isasymGamma}}}} {\sc Skip}
\\[2ex]
\isa{\mbox{}\inferrule{\mbox{E{\isacharcomma}\ M{\isacharcomma}\ {\isasymGamma}\ {\isasymturnstile}\ x\ {\isasymleftarrow}\ {\isacharparenleft}b{\isacharcomma}\ None{\isacharparenright}{\isacharcomma}\ sl}\\\ \mbox{E{\isacharcomma}\ M{\isacharcomma}\ {\isasymGamma}\ {\isasymturnstile}\ e\ {\isasymrightarrow}\ v{\isacharcomma}\ sv}\\\ \mbox{s\ {\isacharequal}\ sl\ {\isasymsqunion}\ sv\ {\isasymsqunion}\ pc}\\\ \mbox{s{\isacharprime}\ {\isacharequal}\ sl\ {\isasymsqunion}\ pc}\\\ \mbox{M{\isacharprime}\ {\isacharequal}\ M{\isacharparenleft}b\ {\isacharcolon}{\isacharequal}\ ScalarVal\ v{\isacharparenright}}\\\ \mbox{{\isasymGamma}{\isacharprime}\ {\isacharequal}\ {\isasymGamma}{\isacharparenleft}b\ {\isacharcolon}{\isacharequal}\ s{\isacharparenright}}\\\ \mbox{{\isasymGamma}{\isacharprime}{\isacharprime}\ {\isacharequal}\ update\ S\isactrlsub P\ {\isacharparenleft}x\ {\isacharcolon}{\isacharcolon}{\isacharequal}\ e{\isacharparenright}\ s{\isacharprime}\ {\isasymGamma}{\isacharprime}}}{\mbox{E{\isacharcomma}\ S\isactrlsub P{\isacharcomma}\ pc\ {\isasymturnstile}\ x\ {\isacharcolon}{\isacharcolon}{\isacharequal}\ e{\isacharcomma}\ M{\isacharcomma}\ {\isasymGamma}\ {\isasymRightarrow}\ M{\isacharprime}{\isacharcomma}\ {\isasymGamma}{\isacharprime}{\isacharprime}}}} {\sc AssignScalar}
\\[2ex]
\isa{\mbox{}\inferrule{\mbox{E{\isacharcomma}\ M{\isacharcomma}\ {\isasymGamma}\ {\isasymturnstile}\ x\ {\isasymleftarrow}\ {\isacharparenleft}b{\isacharcomma}\ Some\ i{\isacharparenright}{\isacharcomma}\ sl}\\\ \mbox{E{\isacharcomma}\ M{\isacharcomma}\ {\isasymGamma}\ {\isasymturnstile}\ e\ {\isasymrightarrow}\ v{\isacharcomma}\ sv}\\\ \mbox{s\ {\isacharequal}\ sl\ {\isasymsqunion}\ sv\ {\isasymsqunion}\ pc}\\\ \mbox{s{\isacharprime}\ {\isacharequal}\ sl\ {\isasymsqunion}\ pc}\\\ \mbox{M\ b\ {\isacharequal}\ ArrayVal\ arr}\\\ \mbox{M{\isacharprime}\ {\isacharequal}\ M{\isacharparenleft}b\ {\isacharcolon}{\isacharequal}\ ArrayVal\ {\isacharparenleft}arr{\isacharparenleft}i\ {\isacharcolon}{\isacharequal}\ v{\isacharparenright}{\isacharparenright}{\isacharparenright}}\\\ \mbox{{\isasymGamma}\ b\ {\isacharequal}\ l}\\\ \mbox{{\isasymGamma}{\isacharprime}\ {\isacharequal}\ {\isasymGamma}{\isacharparenleft}b\ {\isacharcolon}{\isacharequal}\ s\ {\isasymsqunion}\ l{\isacharparenright}}\\\ \mbox{{\isasymGamma}{\isacharprime}{\isacharprime}\ {\isacharequal}\ update\ S\isactrlsub P\ {\isacharparenleft}x\ {\isacharcolon}{\isacharcolon}{\isacharequal}\ e{\isacharparenright}\ s{\isacharprime}\ {\isasymGamma}{\isacharprime}}}{\mbox{E{\isacharcomma}\ S\isactrlsub P{\isacharcomma}\ pc\ {\isasymturnstile}\ x\ {\isacharcolon}{\isacharcolon}{\isacharequal}{\isacharprime}\ e{\isacharcomma}\ M{\isacharcomma}\ {\isasymGamma}\ {\isasymRightarrow}\ M{\isacharprime}{\isacharcomma}\ {\isasymGamma}{\isacharprime}{\isacharprime}}}} {\sc AssignArrayElem}
\\[2ex]
\isa{\mbox{}\inferrule{\mbox{E{\isacharcomma}\ S\isactrlsub P{\isacharcomma}\ pc\ {\isasymturnstile}\ c\isactrlsub {\isadigit{1}}{\isacharcomma}\ M{\isacharcomma}\ {\isasymGamma}\ {\isasymRightarrow}\ M{\isacharprime}{\isacharcomma}\ {\isasymGamma}{\isacharprime}}\\\ \mbox{E{\isacharcomma}\ S\isactrlsub P{\isacharcomma}\ pc\ {\isasymturnstile}\ c\isactrlsub {\isadigit{2}}{\isacharcomma}\ M{\isacharprime}{\isacharcomma}\ {\isasymGamma}{\isacharprime}\ {\isasymRightarrow}\ M{\isacharprime}{\isacharprime}{\isacharcomma}\ {\isasymGamma}{\isacharprime}{\isacharprime}}}{\mbox{E{\isacharcomma}\ S\isactrlsub P{\isacharcomma}\ pc\ {\isasymturnstile}\ c\isactrlsub {\isadigit{1}}\ {\isacharsemicolon}{\isacharsemicolon}\ c\isactrlsub {\isadigit{2}}{\isacharcomma}\ M{\isacharcomma}\ {\isasymGamma}\ {\isasymRightarrow}\ M{\isacharprime}{\isacharprime}{\isacharcomma}\ {\isasymGamma}{\isacharprime}{\isacharprime}}}} {\sc Comp}
\\[2ex]
\isa{\mbox{}\inferrule{\mbox{E{\isacharcomma}\ M{\isacharcomma}\ {\isasymGamma}\ {\isasymturnstile}\ cond\ {\isasymrightarrow}\ v{\isacharcomma}\ s}\\\ \mbox{istrue\ v}\\\ \mbox{pc{\isacharprime}\ {\isacharequal}\ s\ {\isasymsqunion}\ pc}\\\ \mbox{E{\isacharcomma}\ S\isactrlsub P{\isacharcomma}\ pc{\isacharprime}\ {\isasymturnstile}\ then{\isacharunderscore}body{\isacharcomma}\ M{\isacharcomma}\ {\isasymGamma}\ {\isasymRightarrow}\ M{\isacharprime}{\isacharcomma}\ {\isasymGamma}{\isacharprime}}\\\ \mbox{{\isasymGamma}{\isacharprime}{\isacharprime}\ {\isacharequal}\ update\ S\isactrlsub P\ else{\isacharunderscore}body\ pc{\isacharprime}\ {\isasymGamma}{\isacharprime}}}{\mbox{E{\isacharcomma}\ S\isactrlsub P{\isacharcomma}\ pc\ {\isasymturnstile}\ If\ cond\ then{\isacharunderscore}body\ else{\isacharunderscore}body{\isacharcomma}\ M{\isacharcomma}\ {\isasymGamma}\ {\isasymRightarrow}\ M{\isacharprime}{\isacharcomma}\ {\isasymGamma}{\isacharprime}{\isacharprime}}}} {\sc IfT}
\\[2ex]
\isa{\mbox{}\inferrule{\mbox{E{\isacharcomma}\ M{\isacharcomma}\ {\isasymGamma}\ {\isasymturnstile}\ a\ {\isasymrightarrow}\ v{\isacharcomma}\ s}\\\ \mbox{{\isasymnot}\ istrue\ v}\\\ \mbox{pc{\isacharprime}\ {\isacharequal}\ s\ {\isasymsqunion}\ pc}\\\ \mbox{E{\isacharcomma}\ S\isactrlsub P{\isacharcomma}\ pc{\isacharprime}\ {\isasymturnstile}\ else{\isacharunderscore}body{\isacharcomma}\ M{\isacharcomma}\ {\isasymGamma}\ {\isasymRightarrow}\ M{\isacharprime}{\isacharcomma}\ {\isasymGamma}{\isacharprime}}\\\ \mbox{{\isasymGamma}{\isacharprime}{\isacharprime}\ {\isacharequal}\ update\ S\isactrlsub P\ then{\isacharunderscore}body\ pc{\isacharprime}\ {\isasymGamma}{\isacharprime}}}{\mbox{E{\isacharcomma}\ S\isactrlsub P{\isacharcomma}\ pc\ {\isasymturnstile}\ If\ a\ then{\isacharunderscore}body\ else{\isacharunderscore}body{\isacharcomma}\ M{\isacharcomma}\ {\isasymGamma}\ {\isasymRightarrow}\ M{\isacharprime}{\isacharcomma}\ {\isasymGamma}{\isacharprime}{\isacharprime}}}} {\sc IfF}
\\[2ex]
\isa{\mbox{}\inferrule{\mbox{E{\isacharcomma}\ M{\isacharcomma}\ {\isasymGamma}\ {\isasymturnstile}\ cond\ {\isasymrightarrow}\ v{\isacharcomma}\ s}\\\ \mbox{istrue\ v}\\\ \mbox{pc{\isacharprime}\ {\isacharequal}\ s\ {\isasymsqunion}\ pc}\\\ \mbox{E{\isacharcomma}\ S\isactrlsub P{\isacharcomma}\ pc{\isacharprime}\ {\isasymturnstile}\ body{\isacharcomma}\ M{\isacharcomma}\ {\isasymGamma}\ {\isasymRightarrow}\ M{\isacharprime}{\isacharcomma}\ {\isasymGamma}{\isacharprime}}\\\ \mbox{E{\isacharcomma}\ S\isactrlsub P{\isacharcomma}\ pc{\isacharprime}\ {\isasymturnstile}\ While\ cond\ body{\isacharcomma}\ M{\isacharprime}{\isacharcomma}\ {\isasymGamma}{\isacharprime}\ {\isasymRightarrow}\ M{\isacharprime}{\isacharprime}{\isacharcomma}\ {\isasymGamma}{\isacharprime}{\isacharprime}}}{\mbox{E{\isacharcomma}\ S\isactrlsub P{\isacharcomma}\ pc\ {\isasymturnstile}\ While\ cond\ body{\isacharcomma}\ M{\isacharcomma}\ {\isasymGamma}\ {\isasymRightarrow}\ M{\isacharprime}{\isacharprime}{\isacharcomma}\ {\isasymGamma}{\isacharprime}{\isacharprime}}}} {\sc WhileT}
\\[2ex]
\isa{\mbox{}\inferrule{\mbox{E{\isacharcomma}\ M{\isacharcomma}\ {\isasymGamma}\ {\isasymturnstile}\ cond\ {\isasymrightarrow}\ v{\isacharcomma}\ s}\\\ \mbox{{\isasymnot}\ istrue\ v}\\\ \mbox{pc{\isacharprime}\ {\isacharequal}\ s\ {\isasymsqunion}\ pc}\\\ \mbox{{\isasymGamma}{\isacharprime}\ {\isacharequal}\ update\ S\isactrlsub P\ body\ pc{\isacharprime}\ {\isasymGamma}}}{\mbox{E{\isacharcomma}\ S\isactrlsub P{\isacharcomma}\ pc\ {\isasymturnstile}\ While\ cond\ body{\isacharcomma}\ M{\isacharcomma}\ {\isasymGamma}\ {\isasymRightarrow}\ M{\isacharcomma}\ {\isasymGamma}{\isacharprime}}}} {\sc WhileF}
\end{center}%
\end{isamarkuptext}%
\isamarkuptrue%
}
\snip{admissible}{1}{2}{%
\isacommand{fun}\isamarkupfalse%
\ loc{\isacharunderscore}block\ {\isacharcolon}{\isacharcolon}\ {\isachardoublequoteopen}loc\ {\isasymRightarrow}\ block{\isachardoublequoteclose}\ \isakeyword{where}\ {\isachardoublequoteopen}loc{\isacharunderscore}block\ {\isacharparenleft}b{\isacharcomma}\ {\isacharunderscore}{\isacharparenright}\ {\isacharequal}\ b{\isachardoublequoteclose}\isanewline
\isanewline
\isacommand{fun}\isamarkupfalse%
\ admissible\ \isakeyword{where}\isanewline
{\isachardoublequoteopen}admissible\ f\ E\ M\ Skip\ {\isacharequal}\ True{\isachardoublequoteclose}\ {\isacharbar}\isanewline
{\isachardoublequoteopen}admissible\ f\ E\ M\ {\isacharparenleft}x\ {\isacharcolon}{\isacharcolon}{\isacharequal}\ e{\isacharparenright}\ {\isacharequal}\ {\isacharparenleft}{\isasymforall}{\isasymGamma}{\isachardot}{\isasymforall}l{\isachardot}{\isasymforall}s{\isachardot}\ {\isacharparenleft}E{\isacharcomma}\ M{\isacharcomma}\ {\isasymGamma}\ {\isasymturnstile}\ x\ {\isasymleftarrow}\ l{\isacharcomma}\ s{\isacharparenright}\ {\isasymlongrightarrow}\ {\isacharparenleft}loc{\isacharunderscore}block\ l{\isacharparenright}\ {\isasymin}\ f\ x{\isacharparenright}{\isachardoublequoteclose}\ {\isacharbar}\isanewline
{\isachardoublequoteopen}admissible\ f\ E\ M\ {\isacharparenleft}x\ {\isacharcolon}{\isacharcolon}{\isacharequal}{\isacharprime}\ e{\isacharparenright}\ {\isacharequal}\ {\isacharparenleft}{\isasymforall}{\isasymGamma}{\isachardot}{\isasymforall}l{\isachardot}{\isasymforall}s{\isachardot}\ {\isacharparenleft}E{\isacharcomma}\ M{\isacharcomma}\ {\isasymGamma}\ {\isasymturnstile}\ x\ {\isasymleftarrow}\ l{\isacharcomma}\ s{\isacharparenright}\ {\isasymlongrightarrow}\ {\isacharparenleft}loc{\isacharunderscore}block\ l{\isacharparenright}\ {\isasymin}\ f\ x{\isacharparenright}{\isachardoublequoteclose}\ {\isacharbar}\isanewline
{\isachardoublequoteopen}admissible\ f\ E\ M\ {\isacharparenleft}Seq\ a\ b{\isacharparenright}\ {\isacharequal}\isanewline
\ \ \ \ {\isacharparenleft}admissible\ f\ E\ M\ a\ {\isasymand}\isanewline
\ \ \ \ \ \ {\isacharparenleft}{\isasymforall}{\isasymGamma}{\isachardot}{\isasymforall}pc{\isachardot}{\isasymforall}M{\isacharprime}{\isachardot}{\isasymforall}{\isasymGamma}{\isacharprime}{\isachardot}\ {\isacharparenleft}E{\isacharcomma}\ f{\isacharcomma}\ pc\ {\isasymturnstile}\ a{\isacharcomma}\ M{\isacharcomma}\ {\isasymGamma}\ {\isasymRightarrow}\ M{\isacharprime}{\isacharcomma}\ {\isasymGamma}{\isacharprime}{\isacharparenright}\ {\isasymlongrightarrow}\ admissible\ f\ E\ M{\isacharprime}\ b{\isacharparenright}{\isacharparenright}{\isachardoublequoteclose}\ {\isacharbar}\isanewline
{\isachardoublequoteopen}admissible\ f\ E\ M\ {\isacharparenleft}If\ c\ t\ e{\isacharparenright}\ {\isacharequal}\ {\isacharparenleft}admissible\ f\ E\ M\ t\ {\isasymand}\ admissible\ f\ E\ M\ e{\isacharparenright}{\isachardoublequoteclose}\ {\isacharbar}\isanewline
{\isachardoublequoteopen}admissible\ f\ E\ M\ {\isacharparenleft}While\ c\ body{\isacharparenright}\ {\isacharequal}\isanewline
\ \ \ \ {\isacharparenleft}admissible\ f\ E\ M\ body\ {\isasymand}\isanewline
\ \ \ \ \ \ {\isacharparenleft}{\isasymforall}M{\isachardot}{\isasymforall}{\isasymGamma}{\isachardot}{\isasymforall}pc{\isachardot}{\isasymforall}M{\isacharprime}{\isachardot}{\isasymforall}{\isasymGamma}{\isacharprime}{\isachardot}\ admissible\ f\ E\ M\ body\ {\isasymlongrightarrow}\ {\isacharparenleft}E{\isacharcomma}\ f{\isacharcomma}\ pc\ {\isasymturnstile}\ body{\isacharcomma}\ M{\isacharcomma}\ {\isasymGamma}\ {\isasymRightarrow}\ M{\isacharprime}{\isacharcomma}\ {\isasymGamma}{\isacharprime}{\isacharparenright}\ {\isasymlongrightarrow}\isanewline
\ \ \ \ \ \ \ \ admissible\ f\ E\ M{\isacharprime}\ body{\isacharparenright}{\isacharparenright}{\isachardoublequoteclose}%
}
\snip{soundness}{1}{2}{%
\isacommand{theorem}\isamarkupfalse%
\ monitor{\isacharunderscore}soundness{\isacharcolon}\isanewline
\ \ \isakeyword{assumes}\ {\isachardoublequoteopen}E{\isacharcomma}\ S\isactrlsub P{\isacharcomma}\ pc\isactrlsub {\isadigit{1}}\ {\isasymturnstile}\ program{\isacharcomma}\ M\isactrlsub {\isadigit{1}}{\isacharcomma}\ {\isasymGamma}\isactrlsub {\isadigit{1}}\ {\isasymRightarrow}\ M\isactrlsub {\isadigit{1}}{\isacharprime}{\isacharcomma}\ {\isasymGamma}\isactrlsub {\isadigit{1}}{\isacharprime}{\isachardoublequoteclose}\isanewline
\ \ \isakeyword{and}\ \ \ \ \ \ \ {\isachardoublequoteopen}admissible\ S\isactrlsub P\ E\ M\isactrlsub {\isadigit{1}}\ program{\isachardoublequoteclose}\isanewline
\ \ \isakeyword{and}\ \ \ \ \ \ \ {\isachardoublequoteopen}admissible\ S\isactrlsub P\ E\ M\isactrlsub {\isadigit{2}}\ program{\isachardoublequoteclose}\ \ \isanewline
\ \ \isakeyword{and}\ \ \ \ \ \ \ {\isachardoublequoteopen}pc\isactrlsub {\isadigit{2}}\ {\isasymsqsubseteq}\ pc\isactrlsub {\isadigit{1}}{\isachardoublequoteclose}\isanewline
\ \ \isakeyword{and}\ \ \ \ \ \ \ {\isachardoublequoteopen}s\ {\isasymturnstile}\ {\isasymGamma}\isactrlsub {\isadigit{2}}\ {\isasymsqsubseteq}\ {\isasymGamma}\isactrlsub {\isadigit{1}}{\isachardoublequoteclose}\isanewline
\ \ \isakeyword{and}\ \ \ \ \ \ \ {\isachardoublequoteopen}{\isasymGamma}\isactrlsub {\isadigit{1}}{\isacharcomma}\ s\ {\isasymturnstile}\ M\isactrlsub {\isadigit{1}}\ {\isasymsim}\ M\isactrlsub {\isadigit{2}}{\isachardoublequoteclose}\isanewline
\ \ \isakeyword{shows}\ \ \ {\isachardoublequoteopen}{\isacharparenleft}E{\isacharcomma}\ S\isactrlsub P{\isacharcomma}\ pc\isactrlsub {\isadigit{2}}\ {\isasymturnstile}\ program{\isacharcomma}\ M\isactrlsub {\isadigit{2}}{\isacharcomma}\ {\isasymGamma}\isactrlsub {\isadigit{2}}\ {\isasymRightarrow}\ M\isactrlsub {\isadigit{2}}{\isacharprime}{\isacharcomma}\ {\isasymGamma}\isactrlsub {\isadigit{2}}{\isacharprime}{\isacharparenright}\ {\isasymlongrightarrow}\ {\isacharparenleft}s\ {\isasymturnstile}\ {\isasymGamma}\isactrlsub {\isadigit{2}}{\isacharprime}\ {\isasymsqsubseteq}\ {\isasymGamma}\isactrlsub {\isadigit{1}}{\isacharprime}{\isacharparenright}\ {\isasymand}\ {\isacharparenleft}{\isasymGamma}\isactrlsub {\isadigit{1}}{\isacharprime}{\isacharcomma}\ s\ {\isasymturnstile}\ M\isactrlsub {\isadigit{1}}{\isacharprime}\ {\isasymsim}\ M\isactrlsub {\isadigit{2}}{\isacharprime}{\isacharparenright}{\isachardoublequoteclose}%
}
\snip{memorymodel}{1}{2}{%
\isacommand{type{\isacharunderscore}synonym}\isamarkupfalse%
\ loc\ {\isacharequal}\ {\isachardoublequoteopen}block\ {\isacharasterisk}\ int\ option{\isachardoublequoteclose}\ \ \isanewline
\isanewline
\isacommand{datatype}\isamarkupfalse%
\ val\ {\isacharequal}\ Num\ int\ {\isacharbar}\ Ptr\ loc\isanewline
\isanewline
\isacommand{datatype}\isamarkupfalse%
\ block{\isacharunderscore}val\ {\isacharequal}\ ScalarVal\ val\ {\isacharbar}\ ArrayVal\ {\isachardoublequoteopen}{\isacharparenleft}int\ {\isasymRightarrow}\ val{\isacharparenright}{\isachardoublequoteclose}\isanewline
\isanewline
\isacommand{type{\isacharunderscore}synonym}\isamarkupfalse%
\ environment\ {\isacharequal}\ {\isachardoublequoteopen}name\ {\isasymRightarrow}\ block{\isachardoublequoteclose}\ \ \ \ \ \ \ \ \isanewline
\isacommand{type{\isacharunderscore}synonym}\isamarkupfalse%
\ memory\ {\isacharequal}\ {\isachardoublequoteopen}block\ {\isasymRightarrow}\ block{\isacharunderscore}val{\isachardoublequoteclose}\ \ \ \ \ \ \ \ \isanewline
\isacommand{type{\isacharunderscore}synonym}\isamarkupfalse%
\ label{\isacharunderscore}memory\ {\isacharequal}\ {\isachardoublequoteopen}block\ {\isasymRightarrow}\ label{\isachardoublequoteclose}%
}
\snip{exprtypes}{1}{2}{%
\isacommand{datatype}\isamarkupfalse%
\ lval\ {\isacharequal}\ Var\ name\ offs\ {\isacharbar}\ Deref\ expr\isanewline
\isakeyword{and}\ \ \ \ \ \ \ \ \ expr\ {\isacharequal}\ Const\ int\ {\isacharbar}\ Lval\ lval\ {\isacharbar}\ AddrOf\ lval\isanewline
\ \ \ \ \ \ \ \ \ \ \ \ \ \ \ \ \ \ \ \ \ \ \ \ \ \ {\isacharbar}\ BinOp\ expr\ expr\ {\isacharparenleft}\isakeyword{infixl}\ {\isachardoublequoteopen}{\isasymcirc}{\isachardoublequoteclose}\ {\isadigit{5}}{\isadigit{5}}{\isacharparenright}\ {\isacharbar}\ PtrAdd\ expr\ expr\ {\isacharparenleft}\isakeyword{infix}\ {\isachardoublequoteopen}{\isasymoplus}{\isachardoublequoteclose}\ {\isadigit{5}}{\isadigit{4}}{\isacharparenright}\isanewline
\isakeyword{and}\ \ \ \ \ \ \ \ \ offs\ {\isacharequal}\ NoOffset\ {\isacharbar}\ Index\ expr%
}
\snip{evalexpr}{1}{2}{%
\begin{isamarkuptext}%
\begin{center}
\isa{\mbox{}\inferrule{\mbox{E\ x\ {\isacharequal}\ b}\\\ \mbox{E{\isacharcomma}\ M{\isacharcomma}\ {\isasymGamma}\ {\isasymturnstile}\ offs\ {\isasymrightarrow}\isactrlsub o\ offset{\isacharcomma}\ s}}{\mbox{E{\isacharcomma}\ M{\isacharcomma}\ {\isasymGamma}\ {\isasymturnstile}\ Var\ x\ offs\ {\isasymleftarrow}\ {\isacharparenleft}b{\isacharcomma}\ offset{\isacharparenright}{\isacharcomma}\ s}}} {\sc LvalVar}
\\[2ex]
\isa{\mbox{}\inferrule{\mbox{E{\isacharcomma}\ M{\isacharcomma}\ {\isasymGamma}\ {\isasymturnstile}\ a\ {\isasymrightarrow}\ Ptr\ {\isacharparenleft}b{\isacharcomma}\ offs{\isacharparenright}{\isacharcomma}\ s}}{\mbox{E{\isacharcomma}\ M{\isacharcomma}\ {\isasymGamma}\ {\isasymturnstile}\ Deref\ a\ {\isasymleftarrow}\ {\isacharparenleft}b{\isacharcomma}\ offs{\isacharparenright}{\isacharcomma}\ s}}} {\sc LvalMem}
\\[2ex]

\isa{\mbox{}\inferrule{\mbox{}}{\mbox{E{\isacharcomma}\ M{\isacharcomma}\ {\isasymGamma}\ {\isasymturnstile}\ Const\ c\ {\isasymrightarrow}\ Num\ c{\isacharcomma}\ {\isasymbottom}}}} {\sc RvalConst}
\\[2ex]
\isa{\mbox{}\inferrule{\mbox{E{\isacharcomma}\ M{\isacharcomma}\ {\isasymGamma}\ {\isasymturnstile}\ a\ {\isasymleftarrow}\ {\isacharparenleft}b{\isacharcomma}\ None{\isacharparenright}{\isacharcomma}\ sl}\\\ \mbox{M\ b\ {\isacharequal}\ ScalarVal\ v}\\\ \mbox{{\isasymGamma}\ b\ {\isacharequal}\ sr}\\\ \mbox{sl\ {\isasymsqunion}\ sr\ {\isacharequal}\ s}}{\mbox{E{\isacharcomma}\ M{\isacharcomma}\ {\isasymGamma}\ {\isasymturnstile}\ Lval\ a\ {\isasymrightarrow}\ v{\isacharcomma}\ s}}} {\sc RvalScalarLval}
\\[2ex]
\isa{\mbox{}\inferrule{\mbox{E{\isacharcomma}\ M{\isacharcomma}\ {\isasymGamma}\ {\isasymturnstile}\ a\ {\isasymleftarrow}\ {\isacharparenleft}b{\isacharcomma}\ Some\ idx{\isacharparenright}{\isacharcomma}\ sl}\\\ \mbox{M\ b\ {\isacharequal}\ ArrayVal\ arr}\\\ \mbox{arr\ idx\ {\isacharequal}\ v}\\\ \mbox{{\isasymGamma}\ b\ {\isacharequal}\ sr}\\\ \mbox{sl\ {\isasymsqunion}\ sr\ {\isacharequal}\ s}}{\mbox{E{\isacharcomma}\ M{\isacharcomma}\ {\isasymGamma}\ {\isasymturnstile}\ Lval\ a\ {\isasymrightarrow}\ v{\isacharcomma}\ s}}} {\sc RvalArrayLval}
\\[2ex]
\isa{\mbox{}\inferrule{\mbox{E{\isacharcomma}\ M{\isacharcomma}\ {\isasymGamma}\ {\isasymturnstile}\ a\ {\isasymleftarrow}\ l{\isacharcomma}\ s}\\\ \mbox{p\ {\isacharequal}\ Ptr\ l}}{\mbox{E{\isacharcomma}\ M{\isacharcomma}\ {\isasymGamma}\ {\isasymturnstile}\ AddrOf\ a\ {\isasymrightarrow}\ p{\isacharcomma}\ s}}} {\sc RvalRef}
\\[2ex]
\isa{\mbox{}\inferrule{\mbox{E{\isacharcomma}\ M{\isacharcomma}\ {\isasymGamma}\ {\isasymturnstile}\ a\ {\isasymrightarrow}\ Num\ va{\isacharcomma}\ sa}\\\ \mbox{E{\isacharcomma}\ M{\isacharcomma}\ {\isasymGamma}\ {\isasymturnstile}\ b\ {\isasymrightarrow}\ Num\ vb{\isacharcomma}\ sb}\\\ \mbox{eval{\isacharunderscore}binop\ va\ vb\ {\isacharequal}\ v}\\\ \mbox{sa\ {\isasymsqunion}\ sb\ {\isacharequal}\ s}}{\mbox{E{\isacharcomma}\ M{\isacharcomma}\ {\isasymGamma}\ {\isasymturnstile}\ a\ {\isasymcirc}\ b\ {\isasymrightarrow}\ Num\ v{\isacharcomma}\ s}}} {\sc RvalBinop}
\\[2ex]
\isa{\mbox{}\inferrule{\mbox{E{\isacharcomma}\ M{\isacharcomma}\ {\isasymGamma}\ {\isasymturnstile}\ p\ {\isasymrightarrow}\ Ptr\ {\isacharparenleft}b{\isacharcomma}\ Some\ idx{\isacharparenright}{\isacharcomma}\ sb}\\\ \mbox{E{\isacharcomma}\ M{\isacharcomma}\ {\isasymGamma}\ {\isasymturnstile}\ offs\ {\isasymrightarrow}\ Num\ i{\isacharcomma}\ si}\\\ \mbox{l\ {\isacharequal}\ {\isacharparenleft}b{\isacharcomma}\ Some\ {\isacharparenleft}idx\ {\isacharplus}\ i{\isacharparenright}{\isacharparenright}}\\\ \mbox{s\ {\isacharequal}\ sb\ {\isasymsqunion}\ si}}{\mbox{E{\isacharcomma}\ M{\isacharcomma}\ {\isasymGamma}\ {\isasymturnstile}\ p\ {\isasymoplus}\ offs\ {\isasymrightarrow}\ Ptr\ l{\isacharcomma}\ s}}} {\sc RvalPtrAdd}
\\[2ex]

\isa{\mbox{}\inferrule{\mbox{}}{\mbox{E{\isacharcomma}\ M{\isacharcomma}\ {\isasymGamma}\ {\isasymturnstile}\ NoOffset\ {\isasymrightarrow}\isactrlsub o\ None{\isacharcomma}\ {\isasymbottom}}}} {\sc OffsNone}
\isa{\mbox{}\inferrule{\mbox{E{\isacharcomma}\ M{\isacharcomma}\ {\isasymGamma}\ {\isasymturnstile}\ i\ {\isasymrightarrow}\ Num\ idx{\isacharcomma}\ s}}{\mbox{E{\isacharcomma}\ M{\isacharcomma}\ {\isasymGamma}\ {\isasymturnstile}\ Index\ i\ {\isasymrightarrow}\isactrlsub o\ Some\ idx{\isacharcomma}\ s}}} {\sc OffsIdx}
\end{center}%
\end{isamarkuptext}%
\isamarkuptrue%
}
\snip{sequivalencedef}{1}{2}{%
\isacommand{definition}\isamarkupfalse%
\ s{\isacharunderscore}equivalence\ {\isacharcolon}{\isacharcolon}\ {\isachardoublequoteopen}label{\isacharunderscore}memory\ {\isasymRightarrow}\ label\ {\isasymRightarrow}\ memory\ {\isasymRightarrow}\ memory\ {\isasymRightarrow}\ bool{\isachardoublequoteclose}\ {\isacharparenleft}{\isachardoublequoteopen}{\isacharunderscore}{\isacharcomma}\ {\isacharunderscore}\ {\isasymturnstile}\ {\isacharunderscore}\ {\isasymsim}\ {\isacharunderscore}{\isachardoublequoteclose}{\isacharparenright}\ \isakeyword{where}\isanewline
{\isachardoublequoteopen}{\isasymGamma}{\isacharcomma}\ s\ {\isasymturnstile}\ M\isactrlsub {\isadigit{1}}\ {\isasymsim}\ M\isactrlsub {\isadigit{2}}\ \ \ {\isasymequiv}\ \ \ {\isacharparenleft}{\isasymforall}\ b{\isacharcolon}{\isacharcolon}block{\isachardot}\ {\isasymGamma}{\isacharparenleft}b{\isacharparenright}\ {\isasymsqsubseteq}\ s\ {\isasymlongrightarrow}\ mem{\isacharunderscore}equal\ M\isactrlsub {\isadigit{1}}\ M\isactrlsub {\isadigit{2}}\ b{\isacharparenright}{\isachardoublequoteclose}%
}
\snip{sequivalencedefx}{1}{2}{%
\begin{isamarkuptext}%
\begin{isabelle}%
{\isasymGamma}{\isacharcomma}\ s\ {\isasymturnstile}\ M\isactrlsub {\isadigit{1}}\ {\isasymsim}\ M\isactrlsub {\isadigit{2}}\ {\isasymequiv}\ {\isasymforall}b{\isachardot}\ {\isasymGamma}\ b\ {\isasymsqsubseteq}\ s\ {\isasymlongrightarrow}\ mem{\isacharunderscore}equal\ M\isactrlsub {\isadigit{1}}\ M\isactrlsub {\isadigit{2}}\ b%
\end{isabelle}%
\end{isamarkuptext}%
\isamarkuptrue%
}
\snip{lessrestrictiveuptodef}{1}{2}{%
\isacommand{definition}\isamarkupfalse%
\ less{\isacharunderscore}restrictive{\isacharunderscore}up{\isacharunderscore}to\ {\isacharcolon}{\isacharcolon}\ {\isachardoublequoteopen}label\ {\isasymRightarrow}\ label{\isacharunderscore}memory\ {\isasymRightarrow}\ label{\isacharunderscore}memory\ {\isasymRightarrow}\ bool{\isachardoublequoteclose}\ {\isacharparenleft}{\isachardoublequoteopen}{\isacharunderscore}\ {\isasymturnstile}\ {\isacharunderscore}\ {\isasymsqsubseteq}\ {\isacharunderscore}{\isachardoublequoteclose}{\isacharparenright}\ \isakeyword{where}\isanewline
{\isachardoublequoteopen}s\ {\isasymturnstile}\ {\isasymGamma}\isactrlsub {\isadigit{2}}\ {\isasymsqsubseteq}\ {\isasymGamma}\isactrlsub {\isadigit{1}}\ \ \ {\isasymequiv}\ \ \ {\isacharparenleft}{\isasymforall}\ b{\isacharcolon}{\isacharcolon}block{\isachardot}\ {\isasymGamma}\isactrlsub {\isadigit{1}}{\isacharparenleft}b{\isacharparenright}\ {\isasymsqsubseteq}\ s\ {\isasymlongrightarrow}\ {\isasymGamma}\isactrlsub {\isadigit{2}}{\isacharparenleft}b{\isacharparenright}\ {\isasymsqsubseteq}\ {\isasymGamma}\isactrlsub {\isadigit{1}}{\isacharparenleft}b{\isacharparenright}{\isacharparenright}{\isachardoublequoteclose}%
}
\snip{lessrestrictiveuptodefx}{1}{2}{%
\begin{isamarkuptext}%
\begin{isabelle}%
s\ {\isasymturnstile}\ {\isasymGamma}\isactrlsub {\isadigit{2}}\ {\isasymsqsubseteq}\ {\isasymGamma}\isactrlsub {\isadigit{1}}\ {\isasymequiv}\ {\isasymforall}b{\isachardot}\ {\isasymGamma}\isactrlsub {\isadigit{1}}\ b\ {\isasymsqsubseteq}\ s\ {\isasymlongrightarrow}\ {\isasymGamma}\isactrlsub {\isadigit{2}}\ b\ {\isasymsqsubseteq}\ {\isasymGamma}\isactrlsub {\isadigit{1}}\ b%
\end{isabelle}%
\end{isamarkuptext}%
\isamarkuptrue%
}
\snip{exprevalsequiv}{1}{2}{%
\isacommand{lemma}\isamarkupfalse%
\ expr{\isacharunderscore}evaluation{\isacharunderscore}with{\isacharunderscore}s{\isacharunderscore}equivalence{\isacharcolon}\isanewline
\ \ \isakeyword{assumes}\ {\isachardoublequoteopen}s\ {\isasymturnstile}\ {\isasymGamma}\isactrlsub {\isadigit{2}}\ {\isasymsqsubseteq}\ {\isasymGamma}\isactrlsub {\isadigit{1}}{\isachardoublequoteclose}\isanewline
\ \ \isakeyword{and}\ \ \ \ \ \ \ \ \ {\isachardoublequoteopen}{\isasymGamma}\isactrlsub {\isadigit{1}}{\isacharcomma}\ s\ {\isasymturnstile}\ M\isactrlsub {\isadigit{1}}\ {\isasymsim}\ M\isactrlsub {\isadigit{2}}{\isachardoublequoteclose}\isanewline
\ \ \isakeyword{shows}\ \ \ \ \ {\isachardoublequoteopen}{\isasymforall}s\isactrlsub {\isadigit{1}}\ v\isactrlsub {\isadigit{2}}\ s\isactrlsub {\isadigit{2}}{\isachardot}\ s\isactrlsub {\isadigit{1}}\ {\isasymsqsubseteq}\ s\ {\isasymlongrightarrow}\ {\isacharparenleft}E{\isacharcomma}\ M\isactrlsub {\isadigit{1}}{\isacharcomma}\ {\isasymGamma}\isactrlsub {\isadigit{1}}\ {\isasymturnstile}\ a\ {\isasymrightarrow}\ v\isactrlsub {\isadigit{1}}{\isacharcomma}\ s\isactrlsub {\isadigit{1}}{\isacharparenright}\ {\isasymlongrightarrow}\ {\isacharparenleft}E{\isacharcomma}\ M\isactrlsub {\isadigit{2}}{\isacharcomma}\ {\isasymGamma}\isactrlsub {\isadigit{2}}\ {\isasymturnstile}\ a\ {\isasymrightarrow}\ v\isactrlsub {\isadigit{2}}{\isacharcomma}\ s\isactrlsub {\isadigit{2}}{\isacharparenright}\ {\isasymlongrightarrow}\ v\isactrlsub {\isadigit{1}}\ {\isacharequal}\ v\isactrlsub {\isadigit{2}}\ {\isasymand}\ s\isactrlsub {\isadigit{2}}\ {\isasymsqsubseteq}\ s\isactrlsub {\isadigit{1}}{\isachardoublequoteclose}\isanewline
\ \ \isakeyword{and}\ \ \ \ \ \ \ \ \ {\isachardoublequoteopen}{\isasymforall}s\isactrlsub {\isadigit{1}}\ b\isactrlsub {\isadigit{2}}\ s\isactrlsub {\isadigit{2}}{\isachardot}\ s\isactrlsub {\isadigit{1}}\ {\isasymsqsubseteq}\ s\ {\isasymlongrightarrow}\ {\isacharparenleft}E{\isacharcomma}\ M\isactrlsub {\isadigit{1}}{\isacharcomma}\ {\isasymGamma}\isactrlsub {\isadigit{1}}\ {\isasymturnstile}\ b\ {\isasymleftarrow}\ b\isactrlsub {\isadigit{1}}{\isacharcomma}\ s\isactrlsub {\isadigit{1}}{\isacharparenright}\ {\isasymlongrightarrow}\ {\isacharparenleft}E{\isacharcomma}\ M\isactrlsub {\isadigit{2}}{\isacharcomma}\ {\isasymGamma}\isactrlsub {\isadigit{2}}\ {\isasymturnstile}\ b\ {\isasymleftarrow}\ b\isactrlsub {\isadigit{2}}{\isacharcomma}\ s\isactrlsub {\isadigit{2}}{\isacharparenright}\ {\isasymlongrightarrow}\ b\isactrlsub {\isadigit{1}}\ {\isacharequal}\ b\isactrlsub {\isadigit{2}}\ {\isasymand}\ s\isactrlsub {\isadigit{2}}\ {\isasymsqsubseteq}\ s\isactrlsub {\isadigit{1}}{\isachardoublequoteclose}\isanewline
\ \ \isakeyword{and}\ \ \ \ \ \ \ \ \ {\isachardoublequoteopen}{\isasymforall}s\isactrlsub {\isadigit{1}}\ i\isactrlsub {\isadigit{2}}\ s\isactrlsub {\isadigit{2}}{\isachardot}\ s\isactrlsub {\isadigit{1}}\ {\isasymsqsubseteq}\ s\ {\isasymlongrightarrow}\ {\isacharparenleft}E{\isacharcomma}\ M\isactrlsub {\isadigit{1}}{\isacharcomma}\ {\isasymGamma}\isactrlsub {\isadigit{1}}\ {\isasymturnstile}\ c\ {\isasymrightarrow}\isactrlsub o\ i\isactrlsub {\isadigit{1}}{\isacharcomma}\ s\isactrlsub {\isadigit{1}}{\isacharparenright}\ {\isasymlongrightarrow}\ {\isacharparenleft}E{\isacharcomma}\ M\isactrlsub {\isadigit{2}}{\isacharcomma}\ {\isasymGamma}\isactrlsub {\isadigit{2}}\ {\isasymturnstile}\ c\ {\isasymrightarrow}\isactrlsub o\ i\isactrlsub {\isadigit{2}}{\isacharcomma}\ s\isactrlsub {\isadigit{2}}{\isacharparenright}\ {\isasymlongrightarrow}\ i\isactrlsub {\isadigit{1}}\ {\isacharequal}\ i\isactrlsub {\isadigit{2}}\ {\isasymand}\ s\isactrlsub {\isadigit{2}}\ {\isasymsqsubseteq}\ s\isactrlsub {\isadigit{1}}{\isachardoublequoteclose}%
}
\snip{labeltypes}{1}{2}{%
\isacommand{datatype}\isamarkupfalse%
\ type\ {\isacharequal}\ TInt\ {\isacharbar}\ TPtr\ type\ {\isacharbar}\ TArray\ type\ nat\isanewline
\isacommand{datatype}\isamarkupfalse%
\ label{\isacharunderscore}kind\ {\isacharequal}\ Exact\ {\isacharbar}\ Summary\isanewline
\isacommand{datatype}\isamarkupfalse%
\ label{\isacharunderscore}type\ {\isacharequal}\ Label\ label{\isacharunderscore}kind\ nat\ type\isanewline
\isanewline
\isacommand{fun}\isamarkupfalse%
\ ptr{\isacharunderscore}label\ \isakeyword{where}\ {\isachardoublequoteopen}ptr{\isacharunderscore}label\ {\isacharparenleft}Label\ kind\ d\ t{\isacharparenright}\ {\isacharequal}\ Label\ kind\ {\isacharparenleft}d{\isacharplus}{\isadigit{1}}{\isacharparenright}\ {\isacharparenleft}TPtr\ t{\isacharparenright}{\isachardoublequoteclose}\isanewline
\isacommand{fun}\isamarkupfalse%
\ array{\isacharunderscore}label\ \isakeyword{where}\ {\isachardoublequoteopen}array{\isacharunderscore}label\ len\ {\isacharparenleft}Label\ kind\ d\ t{\isacharparenright}\ {\isacharequal}\ Label\ kind\ d\ {\isacharparenleft}TArray\ t\ len{\isacharparenright}{\isachardoublequoteclose}\isanewline
\isanewline
\isacommand{fun}\isamarkupfalse%
\ labels{\isacharunderscore}aux\ \isakeyword{where}\isanewline
{\isachardoublequoteopen}labels{\isacharunderscore}aux\ TInt\ {\isacharequal}\ {\isacharbrackleft}Label\ Exact\ {\isadigit{0}}\ TInt{\isacharbrackright}{\isachardoublequoteclose}\ {\isacharbar}\isanewline
{\isachardoublequoteopen}labels{\isacharunderscore}aux\ {\isacharparenleft}TPtr\ t{\isacharparenright}\ {\isacharequal}\isanewline
\ \ {\isacharbrackleft}Label\ Exact\ {\isadigit{0}}\ TInt{\isacharcomma}\ Label\ Summary\ {\isadigit{1}}\ {\isacharparenleft}TPtr\ TInt{\isacharparenright}{\isacharbrackright}\ {\isacharat}\ map\ ptr{\isacharunderscore}label\ {\isacharparenleft}labels{\isacharunderscore}aux\ t{\isacharparenright}{\isachardoublequoteclose}\ {\isacharbar}\isanewline
{\isachardoublequoteopen}labels{\isacharunderscore}aux\ {\isacharparenleft}TArray\ t\ len{\isacharparenright}\ {\isacharequal}\ map\ {\isacharparenleft}array{\isacharunderscore}label\ len{\isacharparenright}\ {\isacharparenleft}labels{\isacharunderscore}aux\ t{\isacharparenright}{\isachardoublequoteclose}\isanewline
\isanewline
\isacommand{fun}\isamarkupfalse%
\ labels\ \isakeyword{where}\isanewline
{\isachardoublequoteopen}labels\ {\isacharparenleft}TArray\ t\ len{\isacharparenright}\ {\isacharequal}\ {\isacharbrackleft}Label\ Summary\ {\isadigit{0}}\ TInt{\isacharbrackright}\ {\isacharat}\ labels{\isacharunderscore}aux\ {\isacharparenleft}TArray\ t\ len{\isacharparenright}{\isachardoublequoteclose}\ {\isacharbar}\isanewline
{\isachardoublequoteopen}labels\ t\ {\isacharequal}\ labels{\isacharunderscore}aux\ t{\isachardoublequoteclose}%
}

\begin{document}
\maketitle

\begin{abstract}
Information flow analysis checks whether certain pieces of (confidential)
data may affect the results of computations in unwanted ways and thus leak
information. Dynamic information flow analysis adds instrumentation code to
the target software to track flows at run time and raise alarms if a flow
policy is violated; hybrid analyses combine this with preliminary static
analysis.

Using a subset of C as the target language, we extend previous work on
hybrid information flow analysis that handled pointers to scalars. Our
extended formulation handles arrays, pointers to array elements, and pointer
arithmetic. Information flow through arrays of pointers is tracked precisely
while arrays of non-pointer types are summarized efficiently.

A prototype of our approach is implemented using the Frama-C program
analysis and transformation framework. Work on a full machine-checked proof
of the correctness of our approach using Isabelle/HOL is well underway; we
present the existing parts and sketch the rest of the correctness argument.

\end{abstract}

\medskip
\noindent \textbf{Keywords.} information flow, non-interference, termination
insensitive non-interference, hybrid program analysis, formal proof

\section{Motivation}

Information flow analysis is the study of how pieces of confidential data
propagate through programs and affect computations. Typically one wishes to
enforce a security policy stating that confidential data is forbidden from
influencing `public' outputs~\cite{denning.denning-1977}. This concept was
generalized as the \emph{non-interference} property which states that
certain classes of computations must not affect
others~\cite{goguen.meseguer-1982}.

A wide range of non-interference analyses exist, both static and dynamic
ones as well as hybrid combinations. Dynamic analyses are popular because
they are more permissive in general, i.\,e., they reject fewer programs that
are in fact safe, and they allow unsafe programs as long as only safe
program paths are executed. Further, they can be applied to programming
languages such as JavaScript which are not amenable to static analysis and
commonly used in settings where code is loaded dynamically.

For example, web pages can include JavaScript code from several different
servers. Each of these pieces of code can both read and write the entire
document that includes it; this means that confidential personal information
known to one server might be exfiltrated to others. Web browsers can use
dynamic information flow analysis to track the origins of each piece of data
and forbid unwanted flows of possibly sensitive data from one Internet
domain to another~\cite{kerschbaumer-flow,jsflow}.

Not all applications of information flow analysis are directly related to
security or privacy, however; the analysis can also have more general
software engineering uses to enforce application-specific properties. For
example, an industry partner would like us to verify that their code
handling the routing of network packets only depends on packet headers but
not the payload. In terms of information flow analysis, the packet payload
is treated as `confidential' data that is not allowed to affect the handling
of the packet in any way.

In order to be able to enforce such properties, we are developing a hybrid
information flow analysis, trying to unify the best features of static and
dynamic analyses. Our analysis is aimed at a large subset of the C
programming language with the goal of scaling the analysis to real-world
safety-critical C applications. The present paper is a step into this
direction. We are also developing a machine-checked proof of correctness of
the entire approach.

The two main contributions of this paper are thus the following:
\begin{itemize}
\item An extension of a previous hybrid information flow analysis for a
    subset of C that included pointers to scalars; our extension can deal
    with arrays and pointer arithmetic.
\item The formalization of the underlying theory in the Isabelle/HOL proof
    assistant, a full machine-checked proof of the correctness of our
    monitor semantics, and ongoing work on formalizing and proving correct
    our program transformation.
\end{itemize}

The rest of the paper is organized as follows. Section~\ref{sec:examples}
describes our model of information flow and non-interference.
Section~\ref{sec:monitor} describes our fully formalized semantics for
information flow monitoring and its formal proof of soundness.
Section~\ref{sec:transformation} describes the program transformation
implementing the flow monitor for C programs.
Section~\ref{sec:implementation} mentions some features of our concrete
implementation, Section~\ref{sec:related} surveys related work, and
Section~\ref{sec:conclusions} concludes.

\section{Information flow tracking by example}
\label{sec:examples}

We illustrate the problems of information flow tracking with pointers and
arrays in a series of examples. The goal of the information flow analysis is
to ensure that all public outputs of the program are independent of all
secret inputs. That is, running the program twice with the same public
inputs but different secret inputs should give the same public outputs; this
property is called~\emph{non-interference}. In our case, the analysis
dynamically tracks the public/secret status of variables and treats every
variable that is public at the end of the program as an output.

In general, there can be more security levels than just public and secret;
in that case, they are required to form a finite lattice with the bottom
element as the `most public' security level. We assume an attacker who knows
the program's source code and is able to make perfect deductions about
secret inputs from observed public inputs and public outputs. We ignore
timing, nontermination, and other side channels that may also leak secret
information.

For this informal presentation, assume that the variable~\verb|secret| is of
type~\verb|int| and tagged as `secret'. All other variables are initially
public (non-confidential) and of type~\verb|int| unless declared otherwise.
The dynamic part of the analysis described in this paper works by
instrumenting the code with additional monitoring code. Each
variable~\verb|x| is associated with one or more additional~\emph{label
variables} marked here by underlining the variable name and adding optional
suffixes, e.\,g.,~\status{x}. Security levels are tracked as integer
values~0 (public) and~1 (secret). Where levels from different sources must
be taken into account, they are joined using the~\verb!|! (bitwise-or)
operator, ensuring that the result is secret iff one of the inputs is
secret. Most of the examples in this section follow Assaf's
work~\cite{assaf-thesis}.

\begin{figure}
\begin{subfigure}[b]{0.17\textwidth}
\begin{minipage}[t][3.5cm]{\textwidth}
\begin{alltt}
x = secret;\phantom{\{}
z = x + y;
\end{alltt}
\end{minipage}
\caption{Explicit flow}
\label{fig:explicitflow}
\end{subfigure}
\hfill
\begin{subfigure}[b]{0.17\textwidth}
\begin{minipage}[t][3.5cm]{\textwidth}
\begin{alltt}
if (secret) \{
    x = 0;
\} else \{
    y = 1;
\}
\end{alltt}
\end{minipage}
\caption{Implicit flow}
\label{fig:implicitflow}
\end{subfigure}
\hfill
\begin{subfigure}[b]{0.22\textwidth}
\begin{minipage}[t][3.5cm]{\textwidth}
\begin{alltt}
   int *p;\phantom{\{}
   if (secret) \{
       p = &x;
   \} else \{
       p = &y;
   \}
   *p = 1;
\end{alltt}
\end{minipage}
\caption{Pointer-based flow}
\label{fig:pointerflow}
\end{subfigure}
\hfill
\begin{subfigure}[b]{0.3\textwidth}
\begin{minipage}[t][3.5cm]{\textwidth}
\begin{alltt}
int array[2] = \{ 0, 0 \};
array[secret & 1] = 1;
x = array[0];
\end{alltt}
\end{minipage}
\caption{Array-based flow}
\label{fig:arrayflow}
\end{subfigure}
\caption{Examples of the four kinds of information flow handled by our
analysis.}
\label{fig:examples}
\end{figure}

\begin{example}
\label{ex:explicitflow}
The code in Figure~\ref{fig:explicitflow} exhibits \emph{explicit flows} of
secret information from~\verb|secret| to~\verb|x| and then further
to~\verb|z|; the information flows explicitly via assignments. The analysis
must recognize these two variables as secret; their values must not be
output, otherwise some information about~\verb|secret| would leak. These
flows can be monitored by instrumenting the code with the two
assignments~\texttt{\underline{x} = \underline{secret};}
and~\texttt{\underline{z} = \underline{x} | \underline{y};} mirroring the
original assignments.
\end{example}

\begin{example}
\label{ex:implicitflow}
Conditional branches cause \emph{implicit flows} from the condition to any
assignment controlled by the branch. In Figure~\ref{fig:implicitflow}, there
is an implicit flow from~\verb|secret| to~\verb|x| and~\verb|y|: Inspecting
their values may allow an inference whether~\verb|secret| is zero or
nonzero. Implicit flows are tracked by the control context in a
variable~\status{pc} (program counter status), with a new
variant~\status{pc'}, \status{pc''}, \dots for each branching statement. The
initial value of the global~\status{pc} is~0 (public), and every branching
statement's own~\status{pc} variable is computed as the combination of the
directly enclosing~\status{pc} variant and the branch condition's label.
The current~\status{pc} variable must be taken into account for any
assignment.

Additionally, both branches must update the labels of any variables modified
in the~\emph{other} branch to ensure that the flow is captured regardless of
the actual path taken.
\begin{quote}
\begin{alltt}
\status{pc'} = \status{pc} | \status{secret};
if (secret) \{
    x = 0;
    \status{x} = 0 | \status{pc'};
    \status{y} = \status{y} | \status{pc'};
\} else \{
    y = 1;
    \status{y} = 0 | \status{pc'};
    \status{x} = \status{x} | \status{pc'};
\}
\end{alltt}
\end{quote}
Note that constants are public and thus get the label~0. As this is the
neutral element of the~\verb_|_ operator, constants have no influence on the
containing expression's status.
\end{example}

\begin{example}
\label{ex:pointerflow}
An assignment through a pointer introduces a flow from the pointer
expression to every possible pointer target. In our monitor, these targets
are identified by static points-to analysis and updated with the pointer's
label. A label pointer tracks the exact target of the pointer at run time.
This means that a pointer variable~\verb|p| gets~\emph{two} label variables,
\status{p} for the label of the pointer itself and~\status{p\_d1} (of type
pointer to label) for the label of~\verb|p|'s target. Whenever the program
updates~\verb|p| to point to a target~\verb|t|, the monitor
updates~\status{p\_d1} to point to the label~\status{t} of~\verb|t|.

The example in Figure~\ref{fig:pointerflow} is monitored as follows:
\begin{quote}
\begin{alltt}
\status{pc'} = \status{pc} | \status{secret};
if (secret) \{
    p = &x;
    \status{p} = 0 | \status{pc'};
    \status{p\_d1} = &\status{x};
\} else \{
    p = &y;
    \status{p} = 0 | \status{pc'};
    \status{p\_d1} = &\status{y};
\}
*p = 1;
*\status{p\_d1} = 0 | \status{pc};
\status{x} = \status{x} | \status{p} | \status{pc};
\status{y} = \status{y} | \status{p} | \status{pc};
\end{alltt}
\end{quote}
The updates of~\status{x} and~\status{y} are needed to ensure a sound
approximation of the respective labels because it is not known statically
which of the two variables will actually be overwritten.
\end{example}

\begin{example}
\label{ex:arrayflow}
The main contributions of this paper concern the handling of arrays.
Consider the problem of writing to an array at a secret index as in
Figure~\ref{fig:arrayflow}. The analysis must again treat~\verb|x| (and all
of~\verb|array|) as a secret variable: Outputting~\verb|x| or any element
of~\verb|array| at the end of the program would allow an attacker to infer
whether the least significant bit of~\verb|secret| is~1. A single secret
write to an array element thus taints the entire array. For efficient
handling of flows through arrays, we introduce a~\emph{summary label} for
the entire array which is updated monotonically on each write access to the
array.
\begin{quote}
\begin{alltt}
int array[2] = \{ 0, 0 \};
\status{array} = 0;
array[secret & 1] = 1;
\status{array} |= 0 | \status{secret};
x = array[0];
\status{x} = 0 | \status{array};
\end{alltt}
\end{quote}
The~\texttt{|=} operator updates~\status{array} by combining its old value
with the right-hand-side value, i.\,e., it performs the equivalent
of~\texttt{\status{array} = \status{array} | \status{secret}}. Such updates
are monotonic, so an array's summary label can never decrease. As we will
see later, we often also need field-sensitive tracking of array fields in
addition to the summary label.
\end{example}

Note that all examples except the first share a common property: A piece of
code modifies some object, but it is not known statically which one of
several objects (variables or array fields) is affected in a concrete
execution. The dynamic part of the analysis, i.\,e., the instrumentation
code, must be aware of all possible objects that may be affected and update
their statuses to hold a safe over-approximation of the actual status. The
set of possible target objects is computed by a standard points-to analysis,
the static part of our analysis.

\section{Monitor semantics}
\label{sec:monitor}

We can now formalize the intuitive explanations from the previous section as
a system of information flow monitoring semantics of programs.
The semantics compute a~\emph{label memory}~\(\Gamma\) which tracks the
labels of objects in memory. It is then possible to prove that this
semantics satisfies the required non-interference property.

The types, definitions, and proofs described in this section are fully
formalized and checked in the Isabelle/HOL proof assistant~\cite{isabelle}.
Their presentation in the paper was generated automatically, directly from
the Isabelle/HOL formalization. As the full development is~1900 lines long,
we only show some key parts and omit auxiliary definitions,
lemmas, and proofs.\footnote{The entire development is available online
at~\url{http://www.complang.tuwien.ac.at/gergo/tini/}.}

\subsection{Expression semantics}

We formalize a simple imperative language corresponding to a subset of C.
Figure~\ref{fig:memorymodel} shows the basics of the memory model.  The
types \isa{name} and \isa{block} are abstract; the type \isa{label} is
required to be some bounded lattice with a bottom element~\isasymbottom, a
join operation~\isasymsqunion, and a corresponding partial
order~\isasymsqsubseteq.

\begin{figure}
\begin{quote}
\begin{isabelle}
\memorymodel
\end{isabelle}
\end{quote}
\caption{Memory model of our simple programming language.}
\label{fig:memorymodel}
\end{figure}

A location~\isa{loc} is a pair of a~\isa{block} and an optional offset. A
value~\isa{val} is either a number or a pointer to a~\isa{loc}. A block can
hold a value~\isa{block-val}, which is either a scalar~\isa{val} or an array
of unbounded size represented as a function from~\isa{int} to~\isa{val}.
To simplify the first version of our theory, there are no multi-dimensional
arrays: Array elements are scalars of type~\isa{val}. An environment~\isa{E}
maps names to blocks, a memory~\isa{M} maps blocks to block values, and a
label memory~\(\Gamma\) maps blocks to their security labels.

Figure~\ref{fig:exprtypes} shows the abstract syntax of our expressions. The
representation is designed to be as close as reasonably possible to the one
used by Frama-C~\cite{frama-c}, which in turn is based on
CIL~\cite{necula-cil}. There is a distinction between lvalue
expressions~\isa{lval} which evaluate to objects and rvalue
expressions~\isa{expr} which evaluate to values. An~\isa{lval} may be based
on a variable or a dereference expression. There is an auxiliary
type~\isa{offs} for optional offsets into objects, i.\,e., array indexing.
An~\isa{expr} may be a constant, the value of an~\isa{lval} (obtained using
the~\isa{Lval} constructor), the address of an~\isa{lval}, or a binary
operation on~\isa{expr}s.
We have a generic arithmetic operator~\(\circ\) that is intended to work on
numbers and a pointer addition operator~\(\oplus\) for adding a pointer and
an integer. For simplicity, there are no arithmetic comparisons or boolean
operators.

\begin{figure}
\begin{quote}
\begin{isabelle}
\exprtypes
\end{isabelle}
\end{quote}
\caption{Abstract syntax of expressions in our simple programming language.
The~\textbf{infix} annotations in parentheses define syntactic sugar for
some operators.}
\label{fig:exprtypes}
\end{figure}

Figure~\ref{fig:evalexpr} shows the inference rules capturing our definition
of the semantics of expressions. The rules describe both the concrete
semantics, i.\,e., the value computed by an expression, and our monitor
semantics, i.\,e., the security label assigned to the expression. A
judgement~\(E, M, \Gamma \vdash e \rightarrow v, s\) means that in the
context of an environment~\(E\), a memory~\(M\), and a label
memory~\(\Gamma\), the expression~\(e\) evaluates (as an rvalue) to the
value~\(v\) and the security label~\(s\). There are corresponding
relations~\(\leftarrow\) for the evaluation of lvalue expressions to
locations and~\(\rightarrow_o\) for offset expressions to optional integers.
Offsets~\isa{Some i}, which arise from evaluating an~\isa{Index} offset
expression, are used for array locations only. Scalar locations
have offset~\isa{None}, which is the value corresponding to a~\isa{NoOffset}
offset expression.

\begin{figure}[p]
\begin{isabelle}
\evalexpr
\end{isabelle}
\caption{Inference rules defining the semantics of expressions in our
example programming language. Judgements compute both concrete values and
security label values of expressions.}
\label{fig:evalexpr}
\end{figure}

As an example, consider the C expression~\verb|arr[idx]| where~\texttt{arr}
and~\texttt{idx} are variables. It is represented in the abstract syntax
as~\isa{Var \(\mathtt{arr}\) (Index (Lval (Var \(\mathtt{idx}\)
NoOffset)))}. For evaluating it as an lvalue, the~\textsc{LvalVar} rule
applies, and the memory block for~\verb|arr| is determined from the
environment~\(E\). The index expression~\verb|idx| can be evaluated to an
integer offset using the~\textsc{OffsIdx} rule and further recursive rule
applications of~\textsc{LvalVar} and~\textsc{OffsNone}.

Note that this presentation does not include static typing of expressions.
Using scalar values with an index or array values without an index is a type
error, as is interchanging~\isa{Num} and~\isa{Ptr} values. As usual, the
semantics simply gets stuck in such cases. Note also that the generic binary
operator~\(\circ\) is interpreted by some unspecified~\isa{eval-binop}
function whose details we do not care about.

Expressions' security labels are computed by the semantics by merging the
labels of subexpressions using the label lattice's~\(\isasymsqunion\)
operation. Constants and the locations of variables are considered
public~(\(\isasymbottom\)), while the labels of memory locations are
read from the label memory~\(\Gamma\) whenever the value of the memory
location is read from the memory~\(M\) 
in the~\textsc{RvalScalarLval} and~\textsc{RvalArrayLval} rules.

\subsection{Statement semantics}

Figure~\ref{fig:instr} shows the abstract syntax of statements of our target
language. The~\isa{Skip} statement, program sequencing, \isa{If}
and~\isa{While} statements are standard. However, for technical reasons (to
make proofs tractable), we currently use two different forms of the
assignment statement: Plain~\isa{Assign} if a value is written to a scalar
location and~\isa{AssignArrayElem} if a value is written to an array
element.

\begin{figure}
\begin{quote}
\begin{isabelle}
\instr
\end{isabelle}
\end{quote}
\caption{Abstract syntax of statements in the example programming language.}
\label{fig:instr}
\end{figure}

\begin{figure}
\begin{isabelle}
\evalstmt
\end{isabelle}
\caption{Semantics of statements in the example programming language. Both
concrete effects on memory and effects on the security label
memory are tracked.}
\label{fig:evalstmt}
\end{figure}

Figure~\ref{fig:evalstmt} shows the semantics of statements, again
describing both concrete semantics (effects of the program on the memory)
and monitor semantics (effects on the security label memory). The judgements
take the form~\(E, S_P, \mathit{pc} \vdash \mathit{program}, M, \Gamma
\Rightarrow M', \Gamma'\). This means that in the context \(E, S_P,
\mathit{pc}\), the program \(\mathit{program}\) evaluated on a memory~\(M\)
and label memory~\(\Gamma\) terminates with a new memory~\(M'\) and new
label memory~\(\Gamma'\). The meaning of the context element~\(S_P\) will be
explained below. \isa{E} is the environment, and~\isa{pc} is the program
counter label.

The concrete parts of the semantics, capturing the computation of the new
memory~\(M\), are standard. The memory can only be modified by assignment.
In the~\textsc{AssignScalar} rule, the assignment's left-hand side~\(x\) is
evaluated to a location consisting of a memory block~\(b\) without an
offset, i.\,e., a scalar location. The right-hand side~\(e\) is evaluated to
a value~\(v\). The new memory is obtained by updating the value stored at
block~\(b\) in the memory to be~\isa{ScalarVal v}.
The~\textsc{AssignArrayElem} rule is similar but more involved. The
assignment's left-hand side~\(x\) evaluates to memory block~\(b\) with an
integer index~\(i\). The memory~\(M\) must contain an array~\isa{arr} at
block~\(b\). The new memory~\(M'\) is obtained by updating~\isa{arr} at
position~\(i\) and storing this new array at block~\(b\).

The concrete semantics of~\isa{If} statements uses an unspecified 
function~\isa{istrue} of type~\isa{val \isasymRightarrow bool} to select one
of the branches to execute. The concrete semantics of~\isa{While} loops
evaluates the body once if the condition is true, then re-applies
a~\isa{While} inference rule in the new memory configuration. A loop
terminates iff the condition becomes false at some point, in which case
the~\textsc{WhileF} rule applies and performs no further changes to the
memory.

The monitor semantics deserves more detailed explanations. Consider first
the expression~\(\Gamma' = \Gamma(b := s)\) in the~\textsc{AssignScalar}
rule. This updates the security label of the target block~\isa{b} to the new
label~\isa{s}, which is computed from the label~\isa{sl} of the assignment's
left-hand side's location, the label~\isa{sv} of the right-hand side value,
and the current program counter label~\isa{pc}. This captures the direct
information flow as shown in Example~\ref{ex:explicitflow}. After this, the
label memory is updated again; the final monitor is~\isa{\(\Gamma''\) =
update \(S_P\) \((x ::= e)\) \(s'\) \(\Gamma'\)}. This captures
pointer-induced flows as demonstrated in Example~\ref{ex:pointerflow}. If
the lvalue~\(x\) is a pointer expression and may refer to different memory
locations at runtime, the labels of each corresponding memory block must be
updated conservatively. This is done by the~\isa{update} function defined in
Figure~\ref{fig:updatedef}. This function takes an alias analysis
function~\(S_P\), a program fragment, a label~\(s\) and a label
memory~\(\Gamma\). It applies the auxiliary function~\isa{collect-updates}
to find all memory blocks that may be modified by the given program
fragment, then produces a new label memory where the label of every block
possibly modified by the program fragment is joined with the label~\(s\). In
the particular case of the~\textsc{AssignScalar} rule, the program fragment
passed to~\isa{update} is the assignment~\(x ::= e\) itself, which means
that the set of blocks to be updated evaluates to just~\isa{\(S_P\) x}.
Correctness of the update depends on a correctness criterion for the~\(S_P\)
function itself. Our formalization uses a predicate~\isa{admissible \(S_P\)
E M program} (shown in Figure~\ref{fig:admissible}) to express that a static
analysis~\(S_P\) computes a safe overapproximation of points-to sets with
respect to the given program, environment, and starting memory.
An alias function is admissible for a program in a certain configuration if
it captures every assignment's target's correctly and is admissible for all
possible configurations that arise in the evaluation of subprograms.

\begin{figure}
\begin{quote}
\begin{isabelle}
\updatedef
\end{isabelle}
\end{quote}
\caption{Definition of the \isa{update} function used to track the effects
of aliasing and unexecuted program paths on the label memory.}
\label{fig:updatedef}
\end{figure}

\begin{figure}
\begin{quote}
\begin{isabelle}
\admissible
\end{isabelle}
\end{quote}
\caption{Definition of the \isa{admissible} predicate on alias functions.}
\label{fig:admissible}
\end{figure}

The~\textsc{AssignArrayElem} rule is similar to~\textsc{AssignScalar} in its
handling of the label memory. The only difference is in the
computation~\(\Gamma' = \Gamma(b := s \sqcup l)\) where~\(l\) is the memory
block's old security label. This means that an array block's label can only
ever increase monotonically, but never decrease. This behavior corresponds
to the discussion of Example~\ref{ex:arrayflow}.

The inference rules involving control flow also use the~\isa{update}
function to capture implicit flows as discussed in
Example~\ref{ex:implicitflow}. After executing one of the branches of
an~\isa{If} statement, \isa{update} is used to adjust the security labels of
all the memory blocks that may be modified in the~\emph{other} branch. Both
the actual execution of one of the branches and the update of the other
branch are performed using an updated program counter label~\(pc'\).
Similarly, even if a~\isa{While} loop never iterates, the labels of all the
objects that may be modified in its body are updated with~\(pc'\). All this
ensures that implicit flows are correctly captured: The labels of objects
that may be modified under the control of the branch condition are at least
as high as the branch condition's label. If the branch condition is
secret, these objects become secret as well, and no public information
escapes that might allow attackers to infer anything about the
condition.

\subsection{Proof of monitor correctness}

After describing the monitor semantics, we can now proceed to its proof of
correctness. Recall that the goal is to prove non-interference: If a program
is run twice on equivalent public inputs but possibly different secret
inputs, all the public outputs must be the same on both runs. This ensures
that the program's (public) output doesn't allow any inferences about the
secret inputs.

The equivalence of public inputs is formalized in the following definition
of~\emph{s-equivalence}. Two memories~\(M_1\) and~\(M_2\) are equivalent up
to a security label~\(s\) if they have the same contents for every memory
block whose label in a certain security memory~\(\Gamma\) is below~\(s\):

\begin{isabelle}
\sequivalencedef
\end{isabelle}

(The \isa{mem-equal} predicate captures equality of the values stored in
block~\(b\) in both memories. We omit its definition for brevity.) 
Somewhat similarly to~\(s\)-equivalence on memories, we define a
predicate imposing a partial ordering on label memories, saying
that~\(\Gamma_2\) is \emph{less restrictive than~\(\Gamma_1\) up to~\(s\)}
if it respects the~\(\sqsubseteq\) ordering on all blocks whose labels are
below~\(s\):

\begin{isabelle}
\lessrestrictiveuptodef
\end{isabelle}

With these definitions, we can state an important lemma saying that the
evaluation of expressions in~\(s\)-equivalent memories is deterministic in a
certain sense:

\begin{isabelle}
\exprevalsequiv
\end{isabelle}

This lemma expresses that if expression evaluation yields a value with a
label below~\(s\), then evaluating the same expression in
an~\(s\)-equivalent configuration will yield the same value and a smaller
(or equal) label. The proof (omitted here) proceeds by mutual induction on
the semantics of evaluation of the different kinds of expressions.

Our main result is the formal proof of the following soundness theorem:

\begin{isabelle}
\soundness
\end{isabelle}

This theorem shows that running the same program twice in \(s\)-equivalent
memories~\(M_1\) and~\(M_2\) (and corresponding side conditions on the
program counter labels and security memories) preserves~\(s\)-equivalence.
Inspection of memory blocks whose labels are below~\(s\) in~\(\Gamma_1'\)
does not yield any information to an attacker. The result only holds if the
static analysis~\(S_P\) is \isa{admissible} for the given program, i.\,e.,
it safely overapproximates all aliasing in the program when started from a
given memory configuration.

\begin{proof}[Proof sketch]
The proof of the soundness theorem proceeds by rule induction on the
semantics. We will sketch the main idea of the soundness argument for
assignments to scalars and for one branch of the evaluation of the~\isa{If}
statement. In either case, the idea is to show preservation
of~\(s\)-equivalence and the `less restrictive up to' relation by
considering how the value and label of an arbitrary memory block~\(b\) is
modified by the program.

\emph{Rule \textsc{AssignScalar}.} We may assume that there exist
derivations in the semantics showing both \(E, S_P, \mathit{pc}_1 \vdash x
::= e, M_1, \Gamma_1 \Rightarrow M_1', \Gamma_1'\) and \(E, S_P,
\mathit{pc}_2 \vdash x ::= e, M_2, \Gamma_2 \Rightarrow M_2', \Gamma_2'\).
In these derivations, name the memory block referenced by~\(x\) as \(b_1\)
and~\(b_2\) and the label of evaluating~\(x\) as an lvalue as~\(s_1\)
and~\(s_2\), respectively. Assume also that there is some arbitrary memory
block~\isa{b} where~\(\Gamma_1'(b) \sqsubseteq s\), i.\,e., \emph{after} the
assignment the label of~\(b\) is below~\(s\). It suffices to show that
\(\Gamma_2'(b) \sqsubseteq \Gamma_1'(b)\) and \(M_1'(b) = M_2'(b)\).

Making a case distinction, assume first that~\isa{b \(\in\) \(S_P\) x}. This
means that~\(b\) may be modified by this assignment according to the
static analysis. It follows that~\(s_1 \sqsubseteq s\) since otherwise
the~\isa{update} function would have changed~\(b\)'s label such
that~\(\Gamma_1'(b) \sqsubseteq s\) would not hold. Using~\(s_1 \sqsubseteq
s\) we can apply the expression evaluation lemma from above to obtain~\(b_1
= b_2\), i.\,e., the same block is assigned in both executions. Further,
if~\(b_1 = b\), i.\,e., this is indeed the block that is modified by the
assignment, another application of the lemma ensures that the same value is
assigned (showing \(M_1'(b) = M_2'(b)\)) and that the expression's labels in
the two derivation trees respect the~\(\sqsubseteq\) ordering,
establishing~\(\Gamma_2'(b) \sqsubseteq \Gamma_1'(b)\). Otherwise, if~\(b_1
\neq b\), then the memory at~\(b\) is not modified at all, and its label
is updated safely using~\isa{update}, again establishing the intended
results.

Finally, in the other case~\isa{b \(\notin\) \(S_P\) x}. Because~\(S_P\) is
an admissible analysis, it follows that~\(b\) is not modified by this
assignment. Hence the semantics rule modifies neither the memory nor the
label memory, and the result follows directly from the assumptions.

\emph{Rule \textsc{IfT}.} Assume there is a derivation
for~\(E, S_P, \mathit{pc}_1 \vdash \isa{If c then-body else-body},
M_1, \Gamma_1 \Rightarrow M_1', \Gamma_1''\) where the condition~\(c\)
evaluates to a true value with label~\(s_1\). From the evaluation of the true
branch~\isa{then-body} in the starting context with updated program
counter~\(\mathit{pc}_1' = s_1 \sqcup \mathit{pc}_1\) obtain a label
memory~\(\Gamma_1'\) where~\(\Gamma_1'' = \isa{update}\ S_P\
\isa{else-body}\ \mathit{pc}_1'\ \Gamma_1'\). Assume further there is a
derivation showing~\(E, S_P, \mathit{pc}_2 \vdash \isa{If c then-body
else-body}, M_2, \Gamma_2 \Rightarrow M_2', \Gamma_2''\). Note that we do
not assume that this derivation enters the same branch. Fix again a
block~\(b\) with~\(\Gamma_1''(b) \sqsubseteq s\).

Making a case distinction, assume~\(s_1 \sqsubseteq s\). Using the
expression evaluation lemma, the branch condition~\(c\) evaluates to a true
value in the second configuration as well, so the same branch is executed.
The required result follows by induction on the execution
of~\isa{then-body}.

Otherwise, \(s_1 \not\sqsubseteq s\). The two derivations may execute
different branches; we show that the block~\(b\) is not affected by
the~\isa{If} statement at all, so the different executions make no
difference to its value or label. First, we have~\isa{b \(\notin\)
collect-updates \(S_P\) else-body} because otherwise the~\isa{update}
function on~\isa{else-body} would have raised its label such that the
assumption~\(\Gamma_1''(b) \sqsubseteq s\) could not hold. Otherwise,
if~\isa{b \(\in\) collect-updates \(S_P\) then-body} were to hold, then at
some point during the execution of the~\isa{If} statement its label would
have to be raised to at least~\(s_1\), again violating the assumption. Thus
we obtain \(\Gamma_2''(b) \sqsubseteq \Gamma_1''(b)\) and \(M_1'(b) =
M_2'(b)\).

The rules for the other branch of the~\isa{If} and for the~\isa{While}
statement follow similar reasoning. Finally, the proof for evaluation
of~\isa{Skip} is trivial, and the proof for program composition follows
directly from the induction hypothesis for the subprograms.
\end{proof}

The full, completely machine-checked Isabelle/HOL proof of this theorem is
about 600 lines long, plus about~200 lines of proofs of key auxiliary
lemmas. The structure of the proof itself follows the work of
Assaf~\cite{assaf-thesis}, which gives a manually typeset paper proof of a
little more than five pages (without handling arrays). We were able to
reproduce the paper proof mostly faithfully, repairing some typographical
errors and minor glitches along the way. The most important issue was that
Assaf's proof of the assignment rule is too weak: His proof only
shows~\(\Gamma_2'(b) \sqsubseteq s\) (for a block~\(b\) modified by the
assignment) rather than the stronger result~\(\Gamma_2'(b) \sqsubseteq
\Gamma_1'(b) \sqsubseteq s\) needed to establish the goal~\(s \vdash
\Gamma_2' \sqsubseteq \Gamma_1'\). However, it was easy to reuse the
structure of the given proof and strengthen it to prove the necessary
condition.

\section{Program transformation}
\label{sec:transformation}

Given the abstract semantics from the previous section, we now turn to the
question of how to implement the security monitor in practice. We want to
insert monitoring code into a given program that tracks security labels.  At
the end of the execution of the program, the label variable~\status{x} for
each original program variable~\verb|x| should have the same value
as~\(\Gamma(E(\texttt{x}))\) in the monitor semantics. The soundness proof
of the monitor then carries over to the analysis code.

\subsection{Information flow monitoring without pointers}

Without pointers or arrays, inlining the dynamic analysis code is simple:
Whenever a variable~\verb|x| is read or written, we insert appropriate reads
or writes of the corresponding label variable~\status{x}. Additionally, for
every statement affecting control flow, a new program counter status
variable is created and updated as in the monitor semantics in
Figure~\ref{fig:evalstmt}. Additional assignments are inserted to model
the effects of the control flow branch not taken, as with the~\isa{update}
function in the monitor semantics.

The difficulties arise when pointers are used: What is the label variable
corresponding to a pointer dereference expression~\verb|*p|? In the abstract
theory, such expressions evaluate to memory blocks~\(b\) which are used to
access both the memory~\(M\) and the label memory~\(\Gamma\). However, these
memory blocks are not available as first-class objects in C, so we need a
different way of finding the correct label variable to access.

\subsection{Information flow monitoring with pointers to scalars}

The solution for tracking pointers developed by Assaf~\cite{assaf-thesis},
which we follow, is to mirror all pointer structures in the original program
in the information flow monitor. For this purpose, each pointer~\verb|p| of
type~\(T \mathtt{*}^{(n)}\) (i.\,e., that may be dereferenced~\(n\) times)
is associated with~\(n\) label pointers~\(\status{\mathtt{p\_d}1}\), \dots,
\(\status{\mathtt{p\_d}\mbox{\(n\)}}\). The intention is to ensure that at
any point in the program, the
expression~\(\mathtt{*}^{(i)}\status{p\_d\mbox{\(i\)}}\) for all \(1 \leq i
\leq n\) evaluates to the label of~\(\mathtt{*}^{(n)}\mathtt{p}\).

For example, if pointer~\verb|p| is made to point to variable~\verb|x| by an
assignment~\verb|p = &x| in the original program, a corresponding label
pointer variable~\status{p\_d1} is made to point to the label
variable~\status{x} by the inserted assignment~\texttt{\status{p\_d1} =
\&\status{x}}.  Reads and writes through~\verb|*p| can then be mirrored in
the analysis as reads and writes through~\verb|*|\status{p\_d1}.
Assaf gives a formal definition of this transformation and proves that it
preserves the invariant that for all pointers in the program, a
pointer~\verb|p| points to a target~\verb|x| iff the corresponding label
pointer points to the target's label. This allows a proof of the correctness
of the transformation, i.\,e., it establishes that the instrumented program
computes the same security labels as the label memory~\(\Gamma\) in the
underlying semantics.

\subsection{Information flow monitoring with arrays}

We extended the approach described above to handle arrays. Note that the
monitor semantics in Figures~\ref{fig:evalexpr} and~\ref{fig:evalstmt}
assume that the memory block storing an array has a single security label,
not individual labels for individual array elements. The reason for this was
touched on in Example~\ref{ex:arrayflow}: If an array element is written at
a secret index, reading another array element and finding it has a
non-secret label would leak information about the value of the index.

For this reason, we associate each array~\verb|a| with a single label
variable~\status{a} called the~\emph{summary label}. As in
the~\textsc{AssignArrayElem} inference rule in the semantics, every write to
an array element triggers a~\emph{weak update} of this label: The summary
label~\(l\) is not overwritten by the new label~\(s\) (which incorporates
the labels of the index and the value to be written) but with the joined
value~\(s \sqcup l\). As security labels form a lattice, we have~\(s
\sqsubseteq s \sqcup l\) and~\(l \sqsubseteq s \sqcup l\). This means that
over a sequence of assignments to elements of the array with labels~\(s_1,
\ldots, s_n\), the values of the summary label~\(l_1, \ldots, l_n\) always
form an ascending chain with respect to~\(\sqsubseteq\). Furthermore, at any
point, the current~\(l_i\) is a safe overapproximation of all~\(s_1, \ldots,
s_{i-1}\) written so far. Our analysis ensures that the label of any
read from array~\verb|a| incorporates its summary label~\status{a}. This
means that, if at any point in the program a secret value or secret index is
used in an assignment to an element of~\verb|a|, all future reads will be
treated as secret. This property ensures the equivalence of~\status{a} to
the label~\(\Gamma(E(\mathtt{a}))\) and hence the soundness of our 
information flow analysis in this aspect.

The summary field also plays an important role in handling pointers to array
elements as well as pointer arithmetic. Consider the following program
fragment:
\begin{quote}
\begin{alltt}
p = &a[i];
p++;
*p = 42;
\end{alltt}
\end{quote}
This code assigns the address of array element~\verb|a[i]| to
pointer~\verb|p|, increments~\verb|p| to point to the next array element,
then writes to memory through~\verb|p|. This final write affects an element
of the array~\verb|a|, so we must ensure that our analysis
updates the summary label~\status{a} correctly.

To this end we must ensure that a label pointer associated with~\verb|p|
always points to the target's summary label and is not moved by pointer
arithmetic. In the example above, a summary
pointer~\status{p\_summary} must be generated by the analysis and
pointed to the address of~\status{a}. This pointer is not affected by
indexing or pointer arithmetic, i.\,e., it always points to~\status{a}
regardless of the value of the index expression~\verb|i| and regardless of
the pointer increment using~\verb|++|. The assignment through~\verb|*p| can
then be mirrored in the analysis by a weak update
through~\texttt{*\status{p\_summary}}, which results in a weak update
of~\status{a} as required.

In the presence of arrays of pointers, a summary label is not enough,
however: We must additionally track pointer relationships in an
array-field-sensitive way. Consider a slightly modified version of the
example above, where~\verb|a| is now an array of pointers rather than an
array of numbers as before, and~\verb|p| is therefore a pointer to a
pointer:
\begin{quote}
\begin{alltt}
p = &a[i];
p++;
a[i+1] = &x;
**p = y;
\end{alltt}
\end{quote}
Here the final assignment through~\verb|**p| is an assignment to the
variable~\verb|x|, and the dynamic information flow analysis must therefore
be able to execute an appropriate update of its label~\status{x}. Thus there
must be an appropriate label pointer~\status{p\_d2}
where~\texttt{**\status{p\_d2}} is the object~\status{x}.

We achieve this by associating a second label with each array of
pointers~\texttt{a[\(n\)]}: Besides the scalar summary label~\status{a}, we
also use an array of label pointers~\texttt{\status{a\_d1}[\(n\)]}. The
intention is to ensure that if~\texttt{a[\(i\)]} points to~\texttt{x},
then~\texttt{\status{a\_d1}[\(i\)]} points to~\status{x}. In the example
above, we can let~\texttt{**\status{p\_d2}} point
to~\texttt{\status{a\_d1}[i]} initially and then mirror the pointer
arithmetic~\verb|p++|. We arrive at the following fragment of monitoring
code (ignoring summary labels for simplicity):
\begin{quote}
\begin{alltt}
\status{p\_d2} = &\status{a\_d1}[i];
\status{p\_d2}++;
\status{a\_d1}[i+1] = &\status{x};
**\status{p\_d2} = \status{y} | \status{p};
\end{alltt}
\end{quote}
The generated code ensures that at the last assignment, \status{p\_d2}
points to~\texttt{\status{a\_d1}[i+1]}, which in turn points to~\status{x}.
The last assignment thus updates~\status{x} as required.

We can thus summarize the requirements for our analysis: Every
array~\verb|a| needs a summary label~\status{a} and an array of exact
labels~\status{a\_d0}. Every pointer~\verb|p| needs a label~\status{p} for
the pointer itself as well as a summary label
pointer~\status{p\_d1\_summary} to point to~\verb|p|'s target's summary
label and a label pointer~\status{p\_d1} to point to~\verb|p|'s exact
target's label. These rules must be applied recursively for types of nested
pointers or arrays, adjusting the number of possible
dereferences~(\texttt{d}). Figure~\ref{fig:labeltypes} shows how we compute
the list of types and dereferencing levels using the function~\isa{labels}.
The recursive computation is captured in the function~\isa{labels-aux}. The
most subtle issue is that~\isa{labels} must add an outermost summary label
for array types.

\begin{figure}
\begin{quote}
\begin{isabelle}
\labeltypes
\end{isabelle}
\end{quote}
\caption{Computation of label types in the presence of arrays and pointers.}
\label{fig:labeltypes}
\end{figure}

For a C type declaration~\verb|int *b[10]|, encoded as~\isa{TArray (TPtr
TInt) 10}, this system computes the following label types, which our program
transformation turns into the appropriate type declarations:
\begin{center}
\begin{tabular}{ll}
\isa{[Label Summary 0 TInt,} &
\verb|int b_status;| \\
\isa{ Label Exact 0 (TArray TInt 10),} &
\verb|int b_status_d0[10];| \\
\isa{ Label Summary 1 (TArray (TPtr TInt) 10),} &
\verb|int *b_status_d1_summary[10];| \\
\isa{ Label Exact 1 (TArray (TPtr TInt) 10)]} &
\verb|int *b_status_d1[10];| \\
\end{tabular}
\end{center}

The types on the left-hand side were computed by the Isabelle/HOL
function~\isa{labels} in Figure~\ref{fig:labeltypes}, the C type
declarations on the right by the equivalent code in our Frama-C plugin
implementing the program transformation for C programs.

\begin{figure}
\hfill
\begin{minipage}[t]{0.15\textwidth}
\begin{alltt}
p = &a[i];
*p = 42;
p += secret;
*p = 43;
\end{alltt}
\end{minipage}
\hfill
\begin{minipage}[t]{0.4\textwidth}
\begin{alltt}
p = & a[i];
\status{p} = 0 | (\status{i} | \status{pc});
\status{p\_d1\_summary} = & \status{a};
\status{p\_d1} = & \status{a\_d0}[i];
*p = 42;
*\status{p\_d1\_summary} |= 0 | (\status{p} | \status{pc});
*\status{p\_d1} = 0 | (\status{p} | \status{pc});
p += secret;
\status{p} |= \status{secret} | \status{pc};
\status{p\_d1\_summary} = \status{p\_d1\_summary};
\status{p\_d1} += \status{secret};
*p = 43;
*\status{p\_d1\_summary} |= 0 | (\status{p} | \status{pc});
*\status{p\_d1} = 0 | (\status{p} | \status{pc});
\end{alltt}
\end{minipage}
\hfill\mbox{}
\caption{Example of dynamic information flow monitoring with arrays and
pointer arithmetic. The original program (left) is turned into the program
with inlined analysis code (right). Our transformation tool's output
was modified to make status variable names more readable, changing names
like~\texttt{p\_status} to~\status{p}.}
\label{fig:example}
\end{figure}

Putting everything together, Figure~\ref{fig:example} shows another variant
of the examples above and the complete dynamic information flow monitoring
code generated by our system. In the statement performing pointer
arithmetic, we use a variable~\verb|secret| to make the flow more visible:
When the pointer~\verb|p| has been offset by~\verb|secret|, its label is
joined with~\verb|secret|'s label. At the subsequent assignment
to~\verb|*p|, this label is propagated to the target's label.
Observe also
how pointer expressions for summary labels perform weak updates (using
the~\verb/|=/ operator), but the corresponding exact labels receive strong
updates.

The remaining challenge is to complete the formalization of this program
transformation in Isabelle/HOL. The key is a precise statement of the
invariant that whenever a pointer expression~\verb|p| points to a
variable~\verb|x|, the corresponding label pointer expression~\status{p}
points to~\status{x}. We will then show that the assignments inserted by the
program transformation preserve this invariant, which will allow us to
establish a complete soundness proof.


\section{Implementation notes}
\label{sec:implementation}

We have implemented the program transformation sketched above as a plugin in
the modular C analysis and transformation framework Frama-C~\cite{frama-c}.
The current prototype handles programs with arbitrary data structures
composed of arrays, pointers, and~\verb|struct| types. For the alias
analysis~\(S_P\) needed by the transformation, we rely on Frama-C's
built-in Value analysis, which computes both aliases for pointers and
value approximations for numeric variables using intervals and other
domains. The transformation is implemented as a transformation of the
Frama-C AST, which can then be output as C code. At the time of writing,
some details of real-world C programs are not yet handled by the analysis,
which precludes us from giving a detailed experimental evaluation of our
approach. The main missing feature is the treatment of~\verb|goto|
statements. These are inserted by Frama-C's AST normalization for
early~\verb|return| statements inside conditionals and for~\verb|continue|
statements in~\verb|for| loops. We are currently working on finalizing the
handling of the information flows due to this kind of non-compositional
control flow.

Using Frama-C's support for code annotations, we allow security levels of
variables to be specified as~\verb|/*@ public */| or~\verb|/*@ private */|
at the point of declaration. The corresponding label variables are then
initialized accordingly. Labels are tracked as integer values of~0 (public)
and~1 (private) and are efficiently combined using the bitwise-or
operator~\verb/|/. We do not currently support more general lattices;
however, extending the current implementation to lattices that can be
represented as bitvectors (up to 64 bits) is straightforward.

Users may also insert annotations
like~\verb|// @ assert security_status(x) == public;| in their programs.
Such annotations may also occur as function preconditions using Frama-C's
annotation language ACSL; for example, any output function could require its
arguments to be public. This allows users full freedom to specify their
application-specific information flow policies. For example, functions that
may cause information to be written to network sockets (such as the
common~\texttt{send(1)} system call) may have contracts requiring their
inputs to be public. As another example, cryptographic code may be annotated
to ensure that branch conditions are always independent of the cryptographic
keys; otherwise, key-dependent control flow may cause differences in timing
or other side-channels observable by
attackers~\cite{boneh.brumley-2003,genkin.etal-2016}. Without such
annotations, our analysis never reports a policy violation, i.\,e., without
a user-defined policy everything is permitted. As such policies are
inherently application-specific, we want to keep our analysis as general as
possible and do not specialize it for particular flow policies.

Transformed, annotated programs often contain enough information for the
Value analysis to be able to prove such assertions without having to execute
the instrumented program at all. Thus our hybrid analysis combined with the
powerful components of the Frama-C framework can often be used as a powerful
static analysis as well.

\section{Related work}
\label{sec:related}

As mentioned several times throughout the paper, our work is heavily based
on the formulation of information flow monitoring by Assaf et
al.~\cite{assaf-paper, assaf-thesis}. This work only handles pointers to
scalars; we have formalized this theory in Isabelle/HOL, extended it to
handle arrays, and are working on extending it further. Our concrete
implementation of the analysis in Frama-C is also based on the prototype
developed by Assaf.

Besides this prototype, we are aware of two implementations of dynamic
information flow analysis that aspire to handle real-world programs. Both of
these are designed for JavaScript and intended for settings with dynamic
code loading. In contrast, our approach assumes a complete program in a
C-like language on which a static points-to analysis can be run. The
approach by Kerschbaumer et al.~\cite{kerschbaumer-flow} handles arrays, but
the details are not described; the authors only mention that an array may
`consist[\dots]\ of heterogeneously labeled fields'.  This heterogeneous
labeling is something our approach consciously avoids for soundness reasons,
to avoid information leaks through array indices. In our approach, reading
an array element always involves reading the array's summary label (see
Example~\ref{ex:arrayflow}). The authors do not describe any formal or
informal proof of non-interference for their analysis.

The other well-developed analysis for JavaScript is JSFlow~\cite{jsflow}
with its extended hybrid version~\cite{jsflow-hybrid}. Both track the labels
of array elements precisely, but a different notion of non-interference from
ours is used: In this variant, it is not allowed to assign secret values to
locations that previously held public values (the converse, overwriting a
secret value by a public value, is allowed). The monitor aborts the program
if a violation of this policy is detected. In our approach, this would
correspond to adding an assertion to every assignment statement. In
contrast, our approach is more permissive and only uses such constraints at
user-defined program points; as discussed above, our analysis is completely
independent of any specific flow policy. The authors prove non-interference
of both versions of JSFlow.

In the literature, there are various static information flow analyses, often
formulated as flow-sensitive type systems~\cite{volpano.etal-1996,
hunt.sands-2006}, as well as further hybrid static/dynamic analyses that are
somewhat comparable to ours~\cite{leguernic.etal-2006,
russo.sabelfeld-2010}. Arrays are occasionally mentioned in connection with
type systems~\cite{volpano.etal-1996} but, to our knowledge, never for the
systems involving some dynamic monitoring. As our work shows, arrays raise
subtle soundness issues, in particular when combined with pointers and
pointer arithmetic; to our knowledge, we are the first ones to handle these
issues in detail for a C-like language.

The terminology of weak and strong updates is borrowed from pointer
analysis~\cite{chase-etal.1990}.

\section{Conclusions and future work}
\label{sec:conclusions}

We presented a hybrid information flow analysis for the C programming
language with pointers, arrays, and pointer arithmetic. Our analysis is
implemented by instrumentation code that tracks information flows by
managing security labels associated with each object in the program. As in
previous work, pointers to labels mirror pointers to data in the original
program. We extend this to arrays, tracking flows both in a field-sensitive
way and as a safe overapproximation in a separate summary field for each
array. Our analysis is implemented using the Frama-C program analysis and
transformation framework.

A machine-checked proof of the correctness of the monitor semantics was
formalized using Isabelle/HOL. We will also formalize the program
transformation and prove its correctness; adding the required static typing
support to our dynamically typed semantics is ongoing work.

We will further extend this work to handle structures in a field-sensitive
way. We also intend to use pointer analysis information to allow
us to handle type casts between pointer types.

\bibliographystyle{eptcs}
\bibliography{vpt2016}
\end{document}